\documentclass[a4paper, prb, twocolumn, showpacs]{revtex4}

\usepackage{amssymb}
\usepackage{amsmath}
\usepackage{color}
\usepackage{graphics, graphicx}
\usepackage{psfrag}
\usepackage{mathcomp}
\usepackage{subfigure}
\usepackage{soul}
\usepackage{ulem} 

\begin{document}

\author{J. Pohlmann$^1$}
\author{A. Privitera$^{1,2}$}
\author{I. Titvinidze$^{1,3}$}
\author{W. Hofstetter$^1$}
\affiliation{1. Institut f\"ur Theoretische Physik, Goethe-Universit\"at, 60438 Frankfurt am Main, Germany\\
2. Democritos National Simulation Center, Consiglio Nazionale delle Ricerche, Istituto Officina dei Materiali (CNR-IOM) 
and International School for Advanced Studies (SISSA), I-34136 Trieste, Italy\\ 
3.  I. Institut f\"ur Theoretische Physik, Universit\"at Hamburg, 20355 Hamburg,  Germany }


\title{Trions and Dimer Formation in Three-Color Fermions}

\begin{abstract}
We study the problem of three ultracold fermions in different hyperfine states loaded into a lattice with spatial dimension $D=1,2$.
We consider $SU(3)$-symmetric attractive interactions and also eventually include a three-body constraint, 
which mimics the effect of three-body losses in the strong-loss regime. We combine exact diagonalization with the
Lanczos algorithm, and evaluate both the eigenvalues and the eigenstates of the problem. 
In $D=1$, we find that the ground state is always a three-body bound state (trion) for arbitrarily small  interaction, while in $D=2$, due to the 
stronger influence of finite-size effects, we are not able to provide conclusive evidence of the existence of a finite threshold for trion formation. 
Our data are however compatible with a threshold value which vanishes logarithmically with the size of the system. Moreover we are able to identify 
the presence of a fine structure inside the spectrum, which is associated with off-site trionic states. The characterization of these states shows 
that only the long-distance behavior of the eigenstate wavefunctions provides clear-cut signatures about the nature of bound states and that onsite 
observables are not enough to discriminate between them. The inclusion of a three-body constraint due to losses promotes these off-site trions to 
the role of lowest energy states, at least in the strong-coupling regime.
\end{abstract}

\maketitle
 
\section{INTRODUCTION}    
The existence of a threshold for bound-state formation in the two-body problem is one of the basic differences between quantum and classical mechanics \cite{2-body_threshold}. 
It is also well known that the existence of this finite threshold strongly depends on the spatial dimension and on the presence (and quantum statistics) of other particles \cite{Cooper}.

Even more striking consequences arise however when we consider the three-body problem. In this case, according to quantum mechanics,
three-body bound states are favored in suitable parameter regimes with respect to two-body bound states. 

This interplay between two-body and three-body physics leads to the so-called Efimov effect \cite{efimov}. Although its theoretical prediction by V. Efimov dates back to 1970, the incredible 
experimental advances in ultracold gases, which allow to observe and study this physical scenario in dilute atomic clouds \cite{grimm_nature, Zaccanti}, have triggered in recent years a strong 
revival of interest \cite{efimov_review}. 

\begin{figure}[hbtp]
\includegraphics[width=0.4\textwidth]{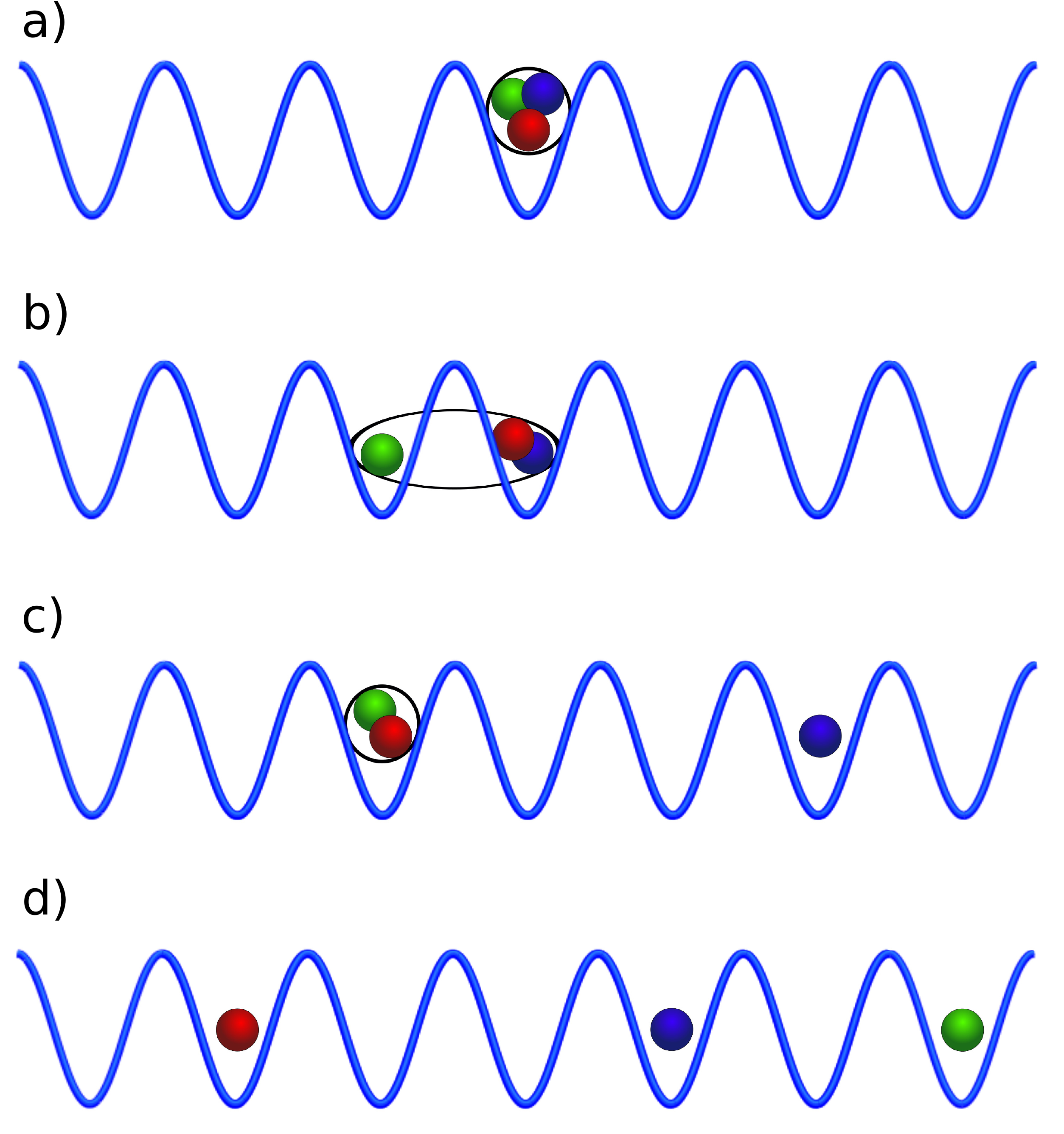}
\caption{(Color online) Sketch of the particle configurations found in the system: (a) onsite trion, (b) off-site trion, (c) dimer + atom and (d) unbound atoms.}
\label{Sketch}
\end{figure}
\begin{figure*}[bt!]
\subfigure[ ]{
\label{GS_wave_function_U0.1}
\begin{minipage}[b]{0.32\textwidth}
\centering \includegraphics[width=1\textwidth]{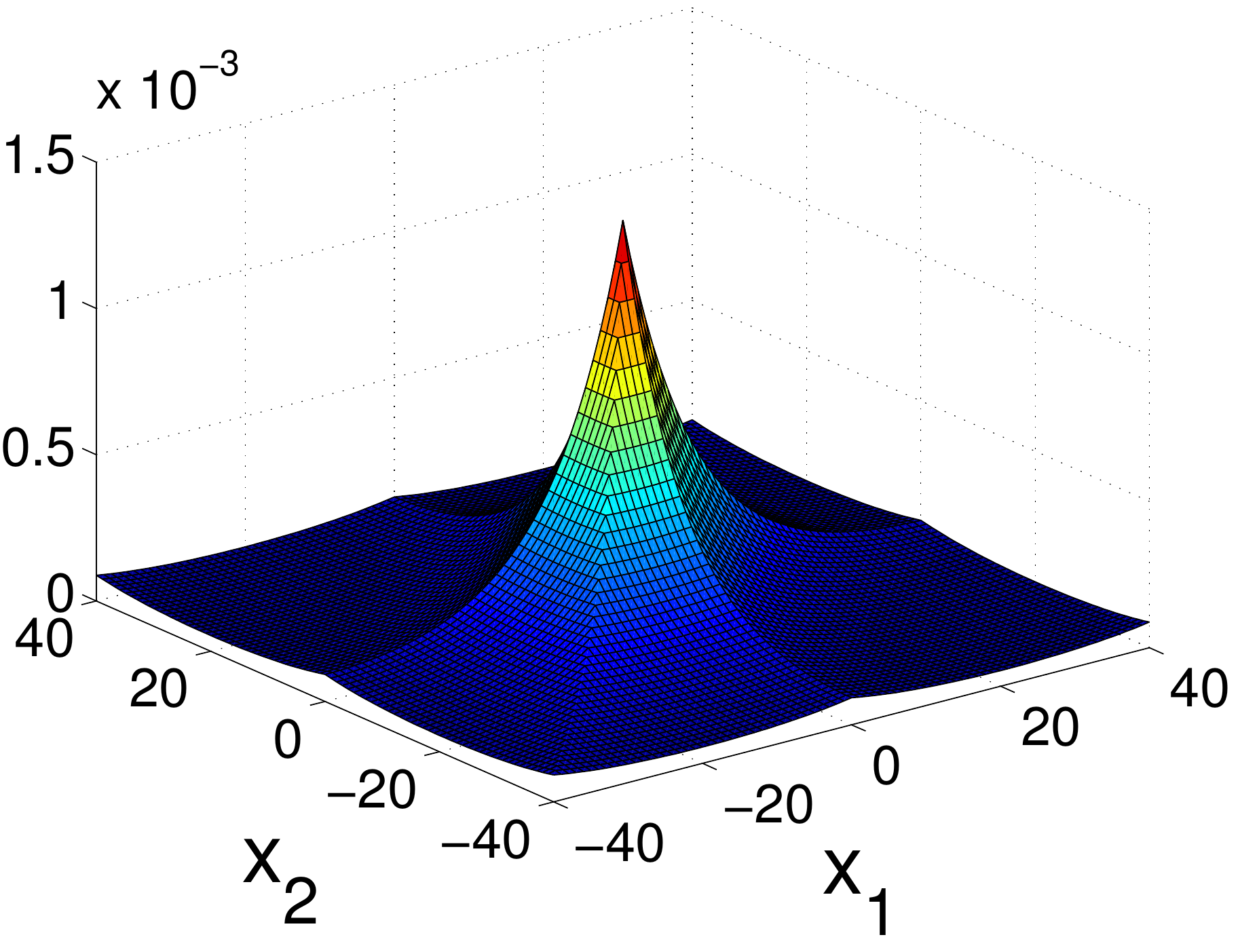}
\end{minipage}}
\subfigure[ ]{
\label{GS_wave_function_U1}
\begin{minipage}[b]{0.32\textwidth}
\centering \includegraphics[width=1\textwidth]{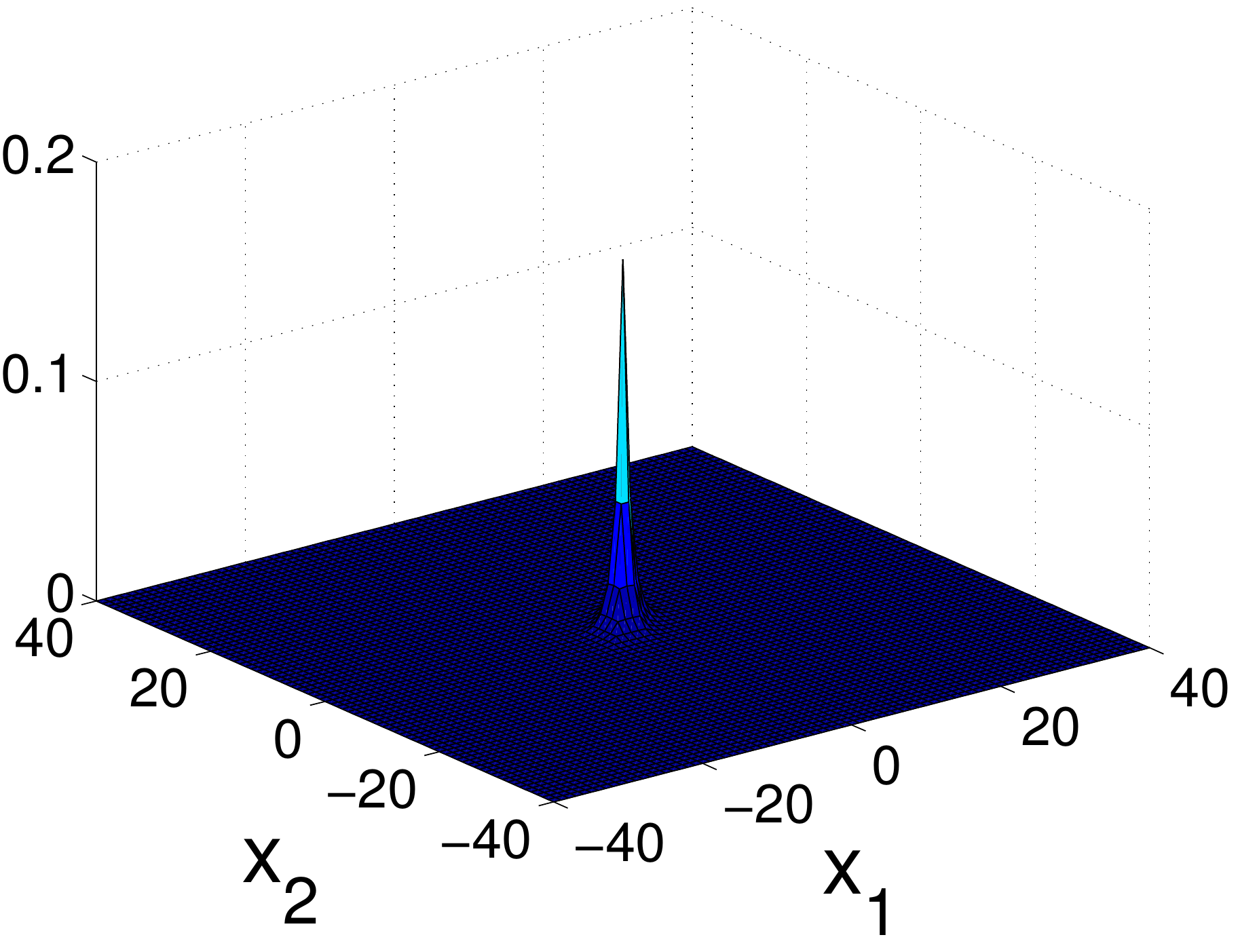}
\end{minipage}}
\subfigure[ ]{
\label{GS_wave_function_U5}
\begin{minipage}[b]{0.32\textwidth}
\centering \includegraphics[width=1\textwidth]{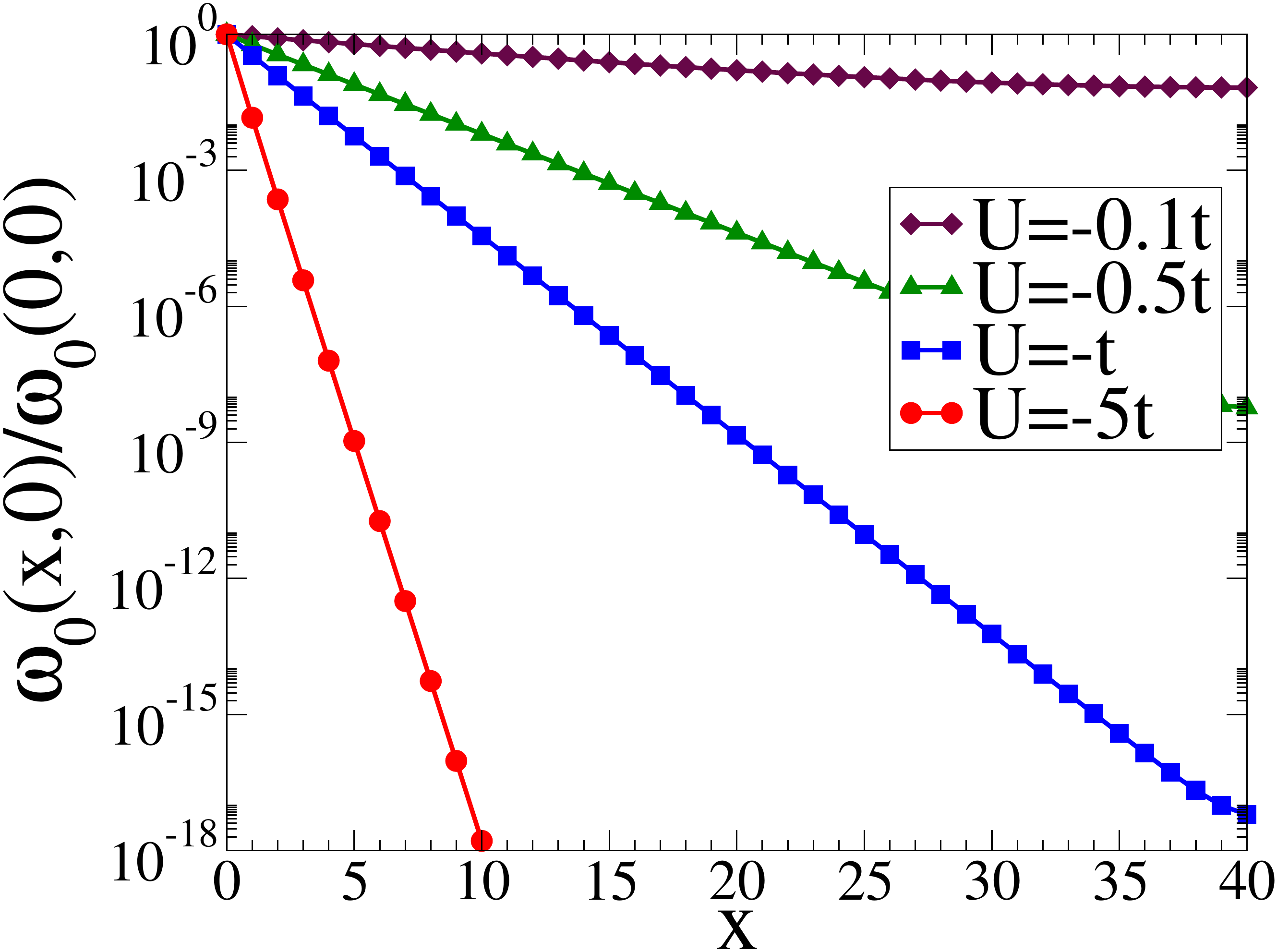}
\end{minipage}}\\
\caption{(Color online) Ground-state probability distribution $\omega_0(x_1,x_2)$ for a linear chain of $N_s=81$ lattice sites for different values of the attraction 
(a) $U/t=-0.1$, and (b) $U/t=-1$. Panel (c) shows the probability distribution for $U=-0.1,-0.5,-1,-5$ plotted in logarithmic scale along the $y=0, x>0$ semi-axis and normalized to the central value,
 in order to underline the exponential decay of the wavefunction.}
\label{GS_wave_function}
\end{figure*}
\begin{figure}[h!]
\subfigure[]{
\label{1D_ns19_spectrum}
\begin{minipage}[b]{0.45\textwidth}
\centering \includegraphics[width=1\textwidth]{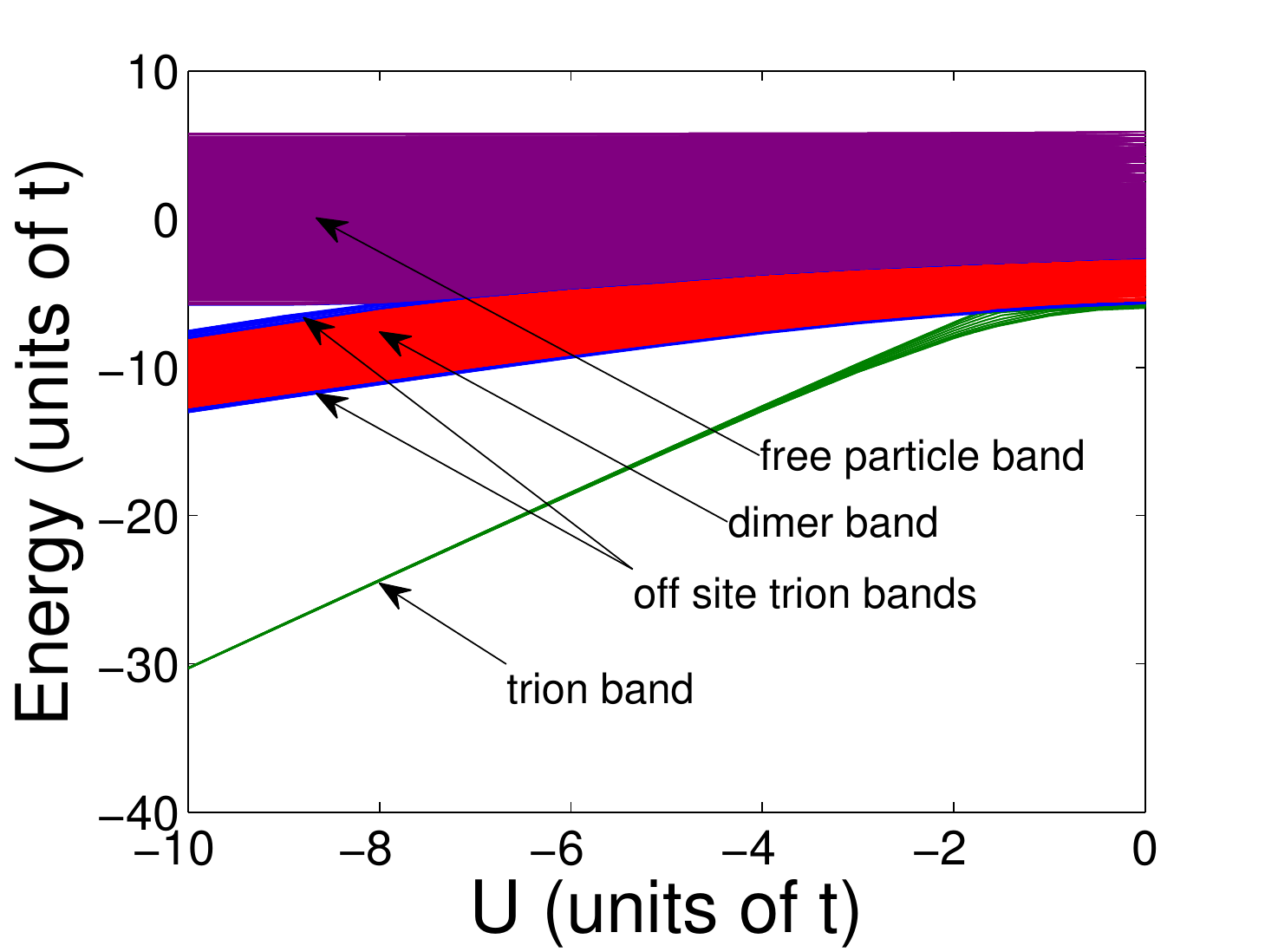}
\end{minipage}}\\
\subfigure[]{
\label{1D_ns51_spectrum}
\begin{minipage}[b]{0.45\textwidth}
\centering \includegraphics[width=1\textwidth]{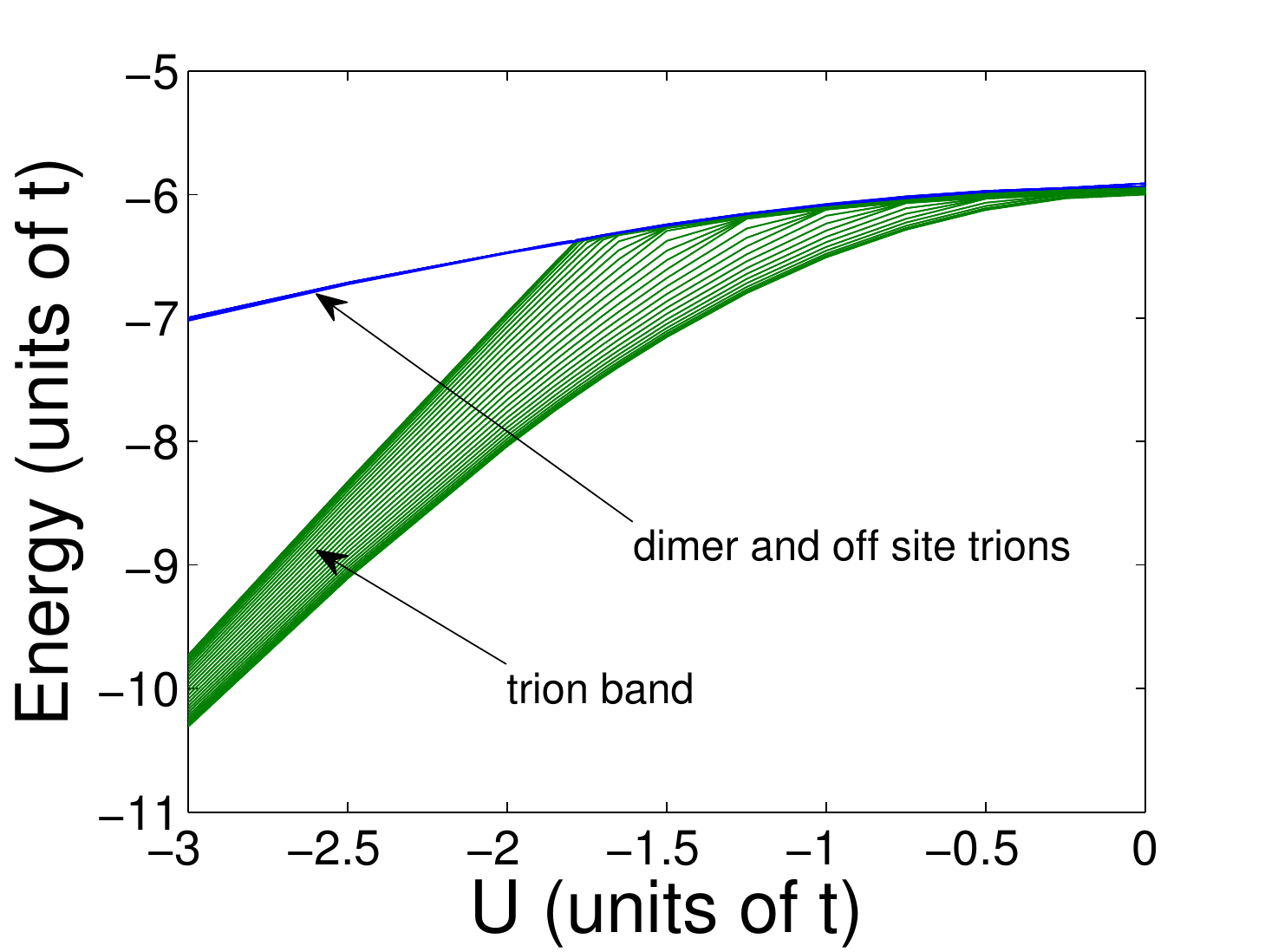}
\end{minipage}}\\
\caption{(Color online) Energy spectrum in 1D as a function of $U/t$. In Panel (a) we plot the full spectrum obtained within ED for  
$N_s=19$ sites, while in Panel (b) we plot the first $N=71$ low-lying eigenvalues obtained within the Lanczos algorithm for $N_s=59$, 
in order to show that the findings in the panel above are not affected by finite-size effects}. 
Different bands are plotted with different color (the trion band is green, the off-site trion band is blue, the dimer band is red and 
the free particle band is purple). For weak interactions bands are not separated and therefore colors are only formal.
\label{Spectrum}
\end{figure}

In this paper we address the problem of three-body bound state formation in a lattice for three fermions with different values of an
internal degree of freedom. Our calculations show that three fermions can form different bound states (see Fig. \ref{Sketch}). 
This is directly relevant for ultracold atomic gases, since the problem under investigation 
corresponds to loading three fermions of the same atomic species (e.g.  $^6{\rm Li}$  [\onlinecite{Ottenstein08, 
Huckans09, Wenz09}] or $^{173}{\rm Yb}$ [\onlinecite{Fukuhara}]) in different hyperfine states into an optical lattice. Similar systems have already been studied 
in current experiments without superimposing an optical lattice, though the experimental observation is hampered by three-body losses 
\cite{Ottenstein08}. Remarkably, by loading the gas into an optical lattice the effect of the losses could be suppressed, as a 
large rate of on-site three-body loss can prevent coherent tunneling processes from populating any site with three particles 
\cite{Daley}. For three-body loss, this mechanism would provide an effective three-body hard-core constraint \cite{diehl, Kantian, Azaria}, 
which suppresses the actual loss events.

The analogous problem in the absence of a lattice has been investigated in Ref. [\onlinecite{Braaten}], and more recently in Ref. [\onlinecite{floerchinger}] by using the functional renormalization 
group, where the authors found a very interesting evolution of the system properties for increasing attraction. Indeed, in the three-dimensional case, three-body bound states appear for smaller values of the attraction than two-body bound states. 
Therefore there is a range of parameters where the ground state is a three-body bound state and no two-body bound states are allowed. In the strong-coupling regime instead a two-body bound state turns out 
to be the stable solution.

This peculiar phenomenon, which is consistent with the Efimov picture, is clearly mirrored 
in the phase diagram of the corresponding many-body problem \cite{floerchinger,baym}, 
though the presence of other fermions triggers pair formation and superfluidity also at weak-coupling.

The presence of a lattice is known to induce important differences in the many-body phase diagram,
especially at strong-coupling and in presence of three-body losses. Indeed, although the overall picture 
of the competition between superfluid and trimer formation (on-site trionic phase) is similar to the continuum case, 
the latter is  stable in the lattice at strong-coupling \cite{Rapp-Hofstetter, Titvinidze-Privitera}, unless the three-body
loss rate $\gamma_3$ is very large \cite{Daley, Titvinidze-Privitera, Privitera-Titvinidze}. In the continuum instead the trionic phase is bound to disappear for 
large values of the attraction \cite{floerchinger,baym}. 

\begin{figure*}[t!]
\subfigure[ ]{
\label{on-site-trions}
\begin{minipage}[b]{0.32\textwidth}
\centering \includegraphics[width=1\textwidth]{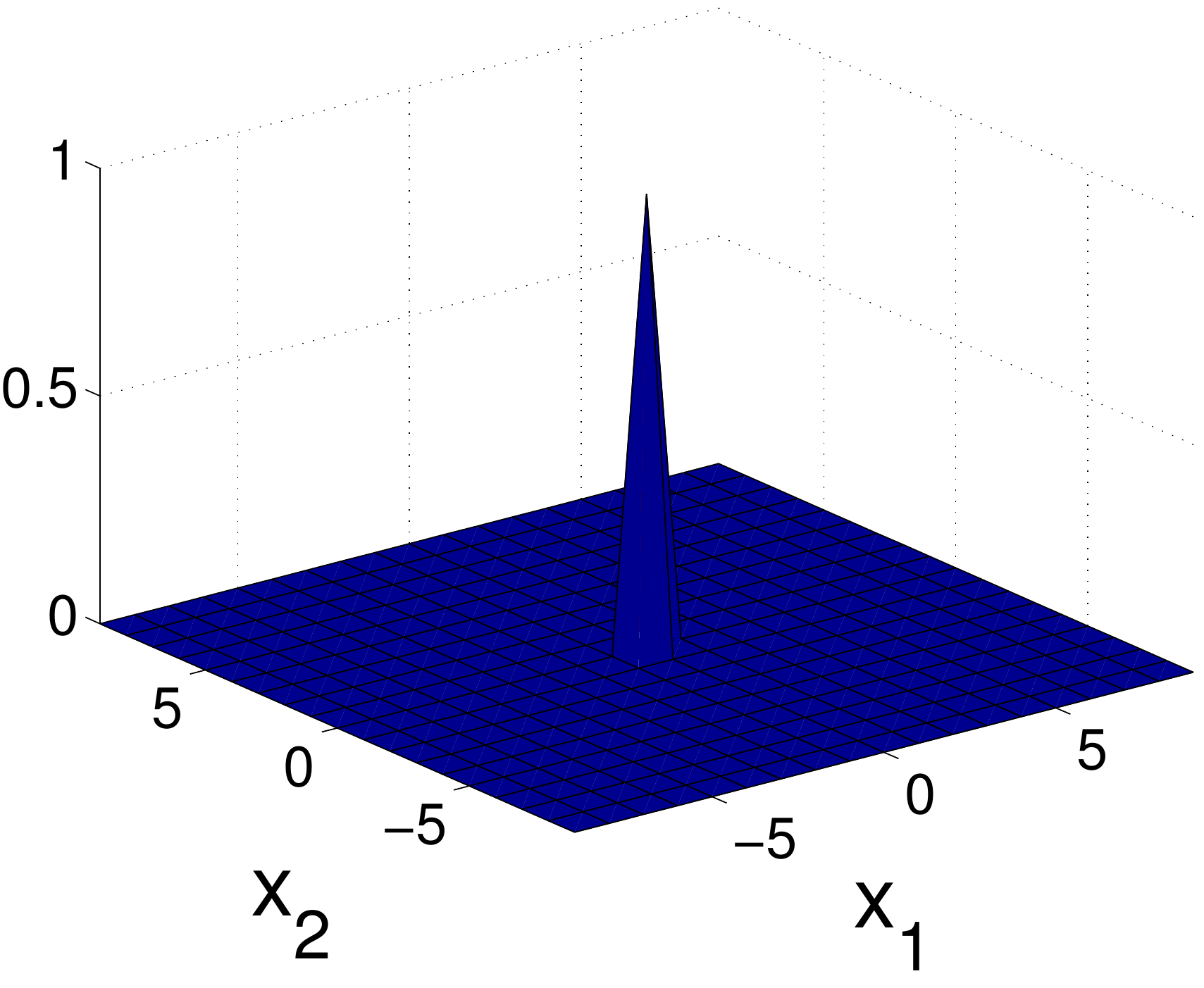}
\end{minipage}}
\subfigure[]{
\label{off-site-trions1}
\begin{minipage}[b]{0.32\textwidth}
\centering \includegraphics[width=1\textwidth]{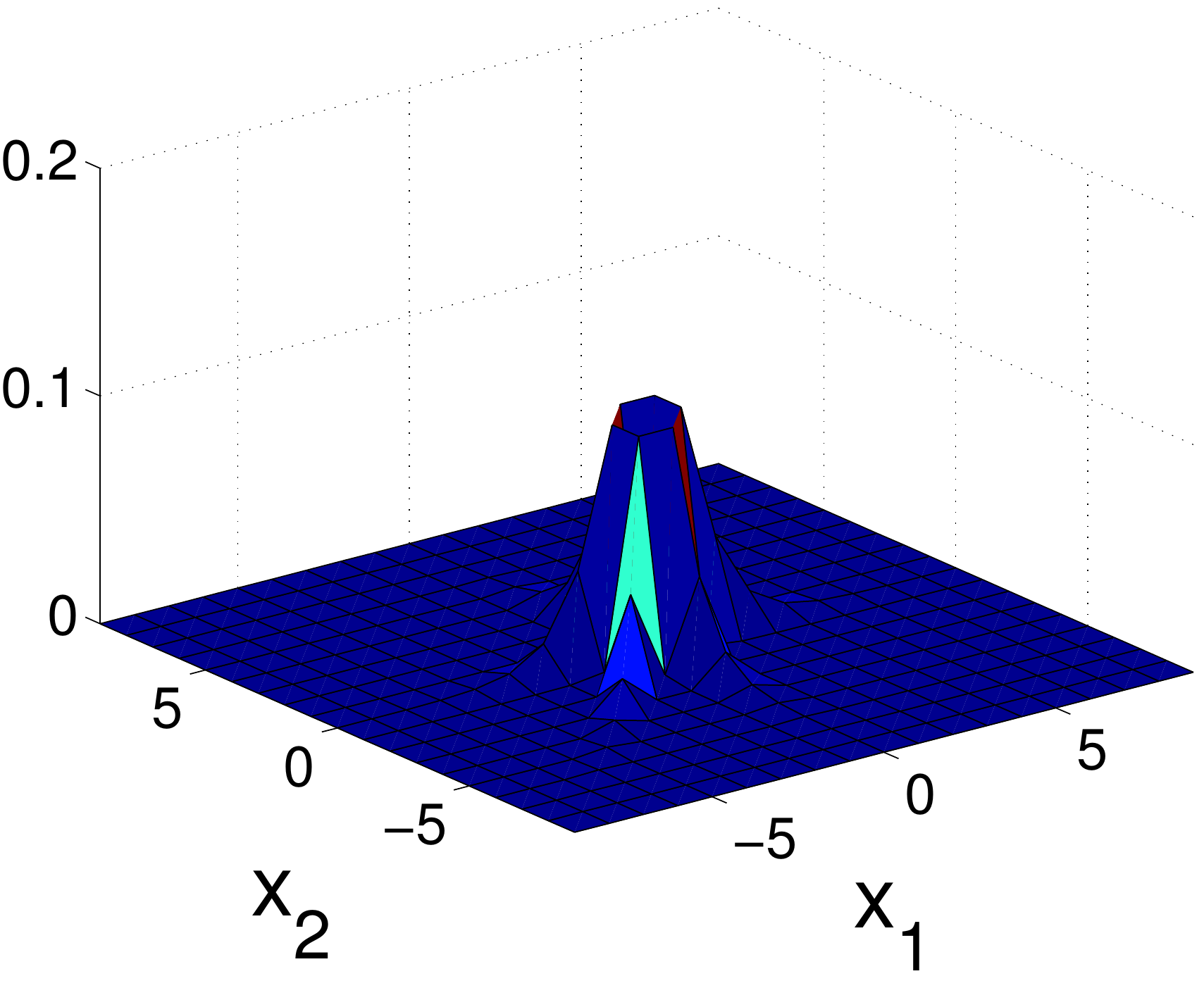}
\end{minipage}}
\subfigure[]{
\label{off-site-trions2}
\begin{minipage}[b]{0.32\textwidth}
\centering \includegraphics[width=1\textwidth]{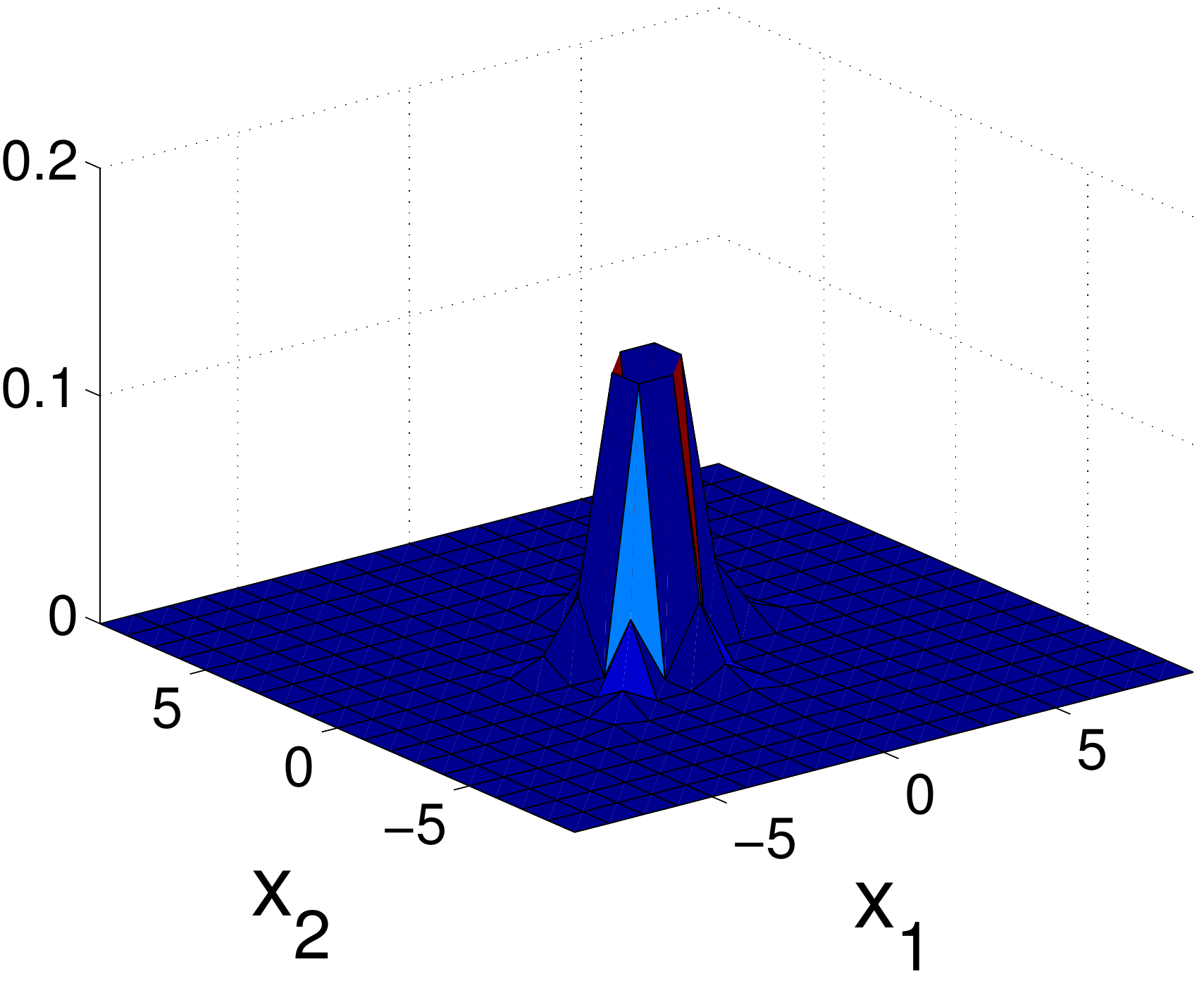}
\end{minipage}}\\
\subfigure[ ]{
\label{dimer1}
\begin{minipage}[b]{0.32\textwidth}
\centering \includegraphics[width=1\textwidth]{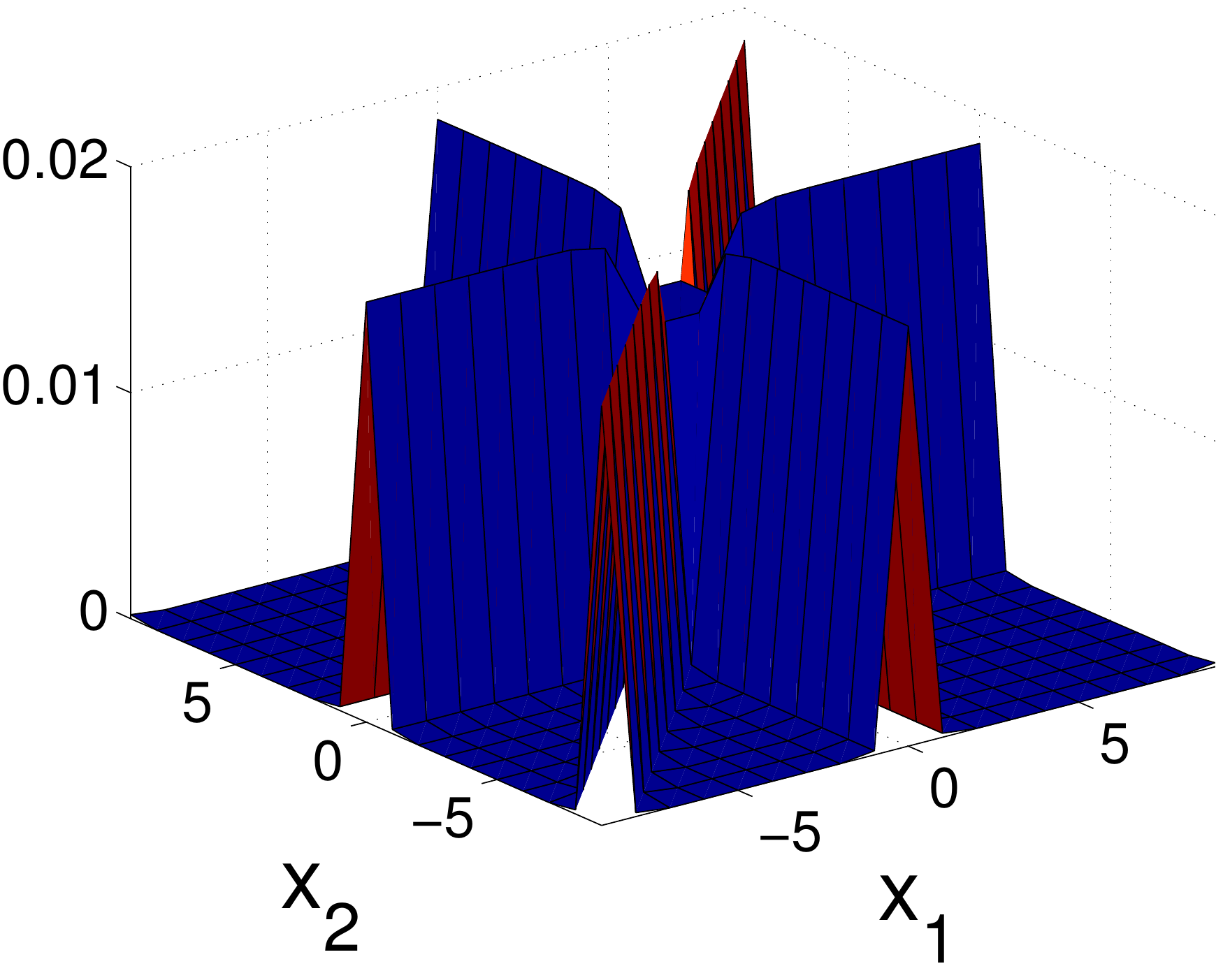}
\end{minipage}}
\subfigure[]{
\label{dimer2}
\begin{minipage}[b]{0.32\textwidth}
\centering \includegraphics[width=1\textwidth]{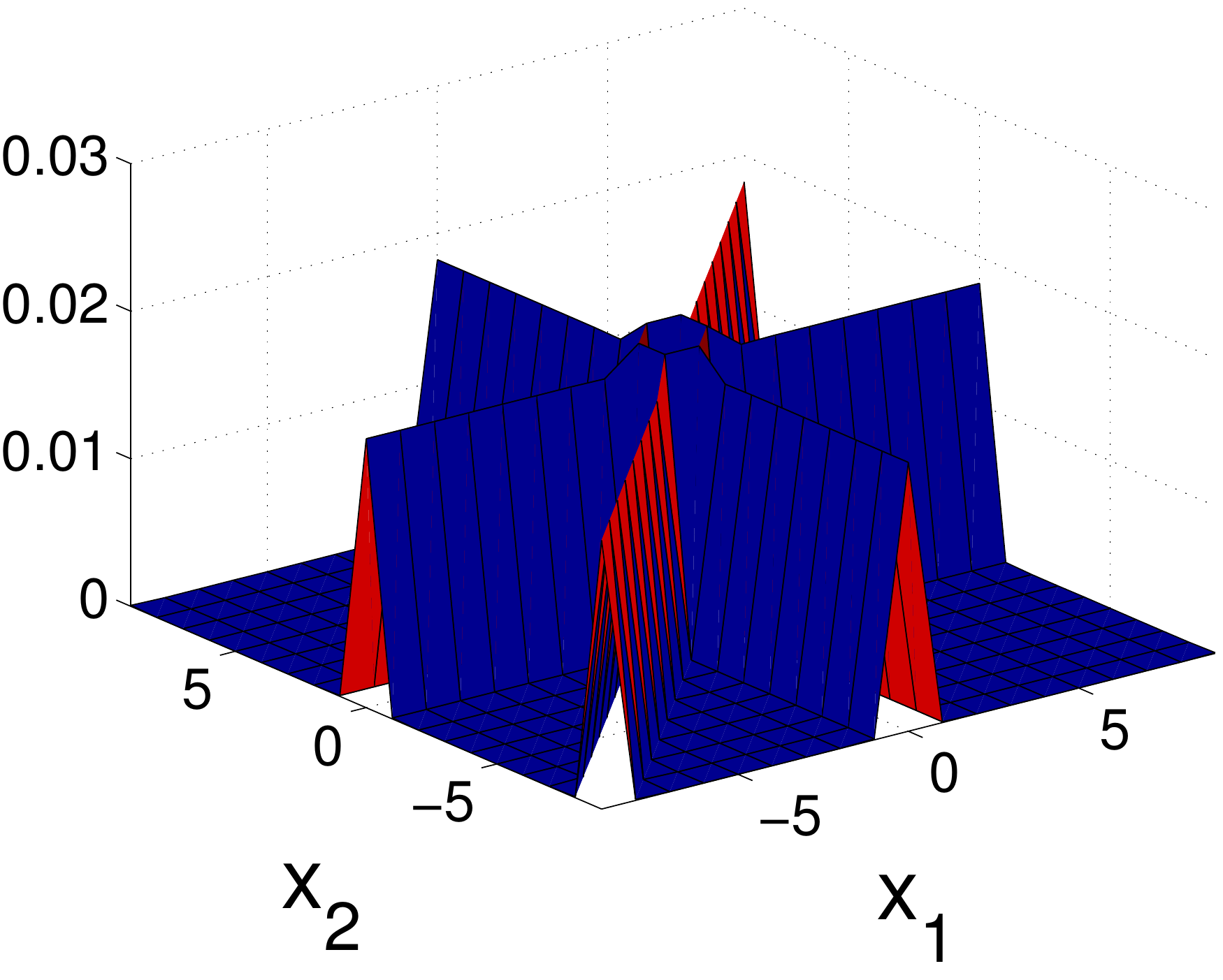}
\end{minipage}}
\subfigure[]{
\label{particle-band}
\begin{minipage}[b]{0.32\textwidth}
\centering \includegraphics[width=1\textwidth]{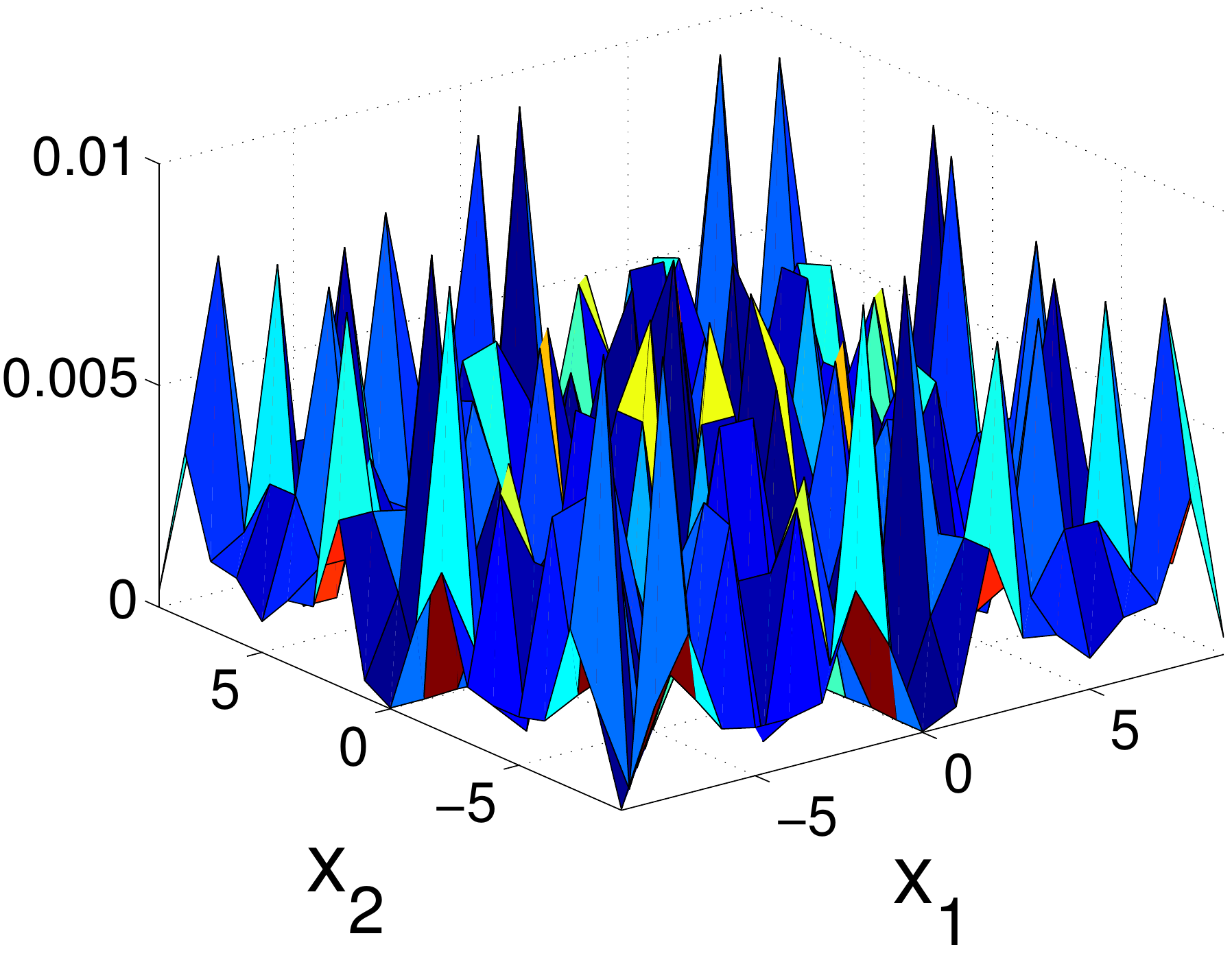}
\end{minipage}}\\
\caption{(Color online) Probability distribution $\omega_n(x_1,x_2)$ corresponding to several excited eigenstates  for $N_s=19$ and $U/t=-20$: (a) excited on-site trion, 
(b) lowest energy off-site trion, (c) highest energy off-site trion from the upper off-site trionic sub-band, (d) lowest energy and (e) excited dimer states, and (f) unbound particles state. 
The large value of the attraction is chosen to be deep in the strong-coupling regime where the band structures are well separated.}
\label{excited_wave_function}
\end{figure*}

In order to better understand this point and the differences to the continuum case, in this paper we take a step back 
and focus on the three-body version of the problem investigated in Ref. [\onlinecite{Titvinidze-Privitera}]. Specifically, we consider
the case where all the interactions between the species are symmetric in a one-dimensional chain ($1D$) and a 
two-dimensional square lattice ($2D$).  We address this problem using a numerical approach, combining exact-diagonalization (ED) with the Lanczos algorithm \cite{matrix_comp}, 
which allows us to obtain information about the low-energy sector of the spectrum
for much larger system sizes than the full ED approach. Moreover we also study the effect of including an onsite three-body hard-core 
constraint, which mimics the effect of a very large three-body loss rate $\gamma_3$ in presence of a lattice, as explained in Ref. [\onlinecite{Daley}]. 

The paper is organized as follows. In the next section, we introduce the model and give details about the methods. 
Then, in Section \ref{sec:1D}, we discuss the results for a one-dimensional chain,  first considering the case without losses, 
then introducing the three-body constraint induced by losses. In Section \ref{sec:2D} we consider the same problem 
in the two-dimensional case, in order to understand what is the role of dimensionality in the formation of three-body bound states and in the competition 
between two- and three-body bound states. Conclusions are drawn in Section \ref{sec:Conclusion}.

\section{Model and Method}

The system under investigation is described by the following single-band Hubbard Hamiltonian:
\begin{equation}
\label{Hamiltonian}
\hat {\cal H} = -t \sum_{\langle ij \rangle, \sigma} \hat c_{i\sigma}^{\dagger}\hat c_{j\sigma}^{\phantom\dagger}+
\frac{U}{2}\sum_{i,\sigma\not=\sigma^\prime}\hat n_{i\sigma}\hat n_{i\sigma^\prime} +
V \sum_{i} \hat n_{i,1}\hat n_{i,2}\hat n_{i,3} \, ,
\end{equation}
where $\langle ij \rangle$ denotes a summation over neighboring lattice sites, $\hat c_{i\sigma}^\dagger$ is the fermionic creation operator with 
hyperfine spin $\sigma=1,2,3$ at site $i$, while $\hat n_{i\sigma}=\hat c_{i\sigma}^{\dagger}\hat c_{i\sigma}^{\phantom\dagger}$ is the 
corresponding number operator. $t$ is the hopping amplitude between nearest neighbor sites and assumed as energy unit, while $U<0$ is the on-site attractive interaction.  
 
In the context of ultracold gases this Hamiltonian can be realized by loading three fermions in different hyperfine states into an optical lattice. 
Under suitable conditions \cite{Titvinidze-Privitera}, the effect of higher lattice bands can be neglected. The case in which the interactions are
symmetric corresponds to equal scattering lengths between different hyperfine states. In practice this can be realized by 
different hyperfine states of $^6{\rm Li}$ in the presence of a strong magnetic field, or by alkaline-earth atoms such as $^{173}{\rm Yb}$.
The Hamiltonian is then (approximately) invariant under $SU(3)$ rotations in the space of the hyperfine states.

The three-body interaction term with $V \to \infty$ (hard-core constraint) is introduced to take the effects of three-body losses in the strong loss regime into account according to 
Refs. [\onlinecite{Daley,Kantian,Azaria}]. On the other hand $V=0$ corresponds to the case when three-body losses are negligible. A finite value of $V$ however, is not equivalent to the regime of moderate loss, 
where actual losses cannot be avoided, and the effective Hamiltonian description above fails. Therefore we do not consider this situation here. 

By using an exact diagonalization approach we were able to calculate 
the full spectrum of the problem for system with up to $N_s=21$ lattice sites for the linear chain in $1D$ and $N_s=16\ (4 \times 4)$ lattice sites for a two-dimensional 
square lattice. Since, however, the size of the system addressed is relatively small (especially for the $2D$ case where the finite-size 
effects are expected to be more severe),  we also used the Lanczos algorithm\cite{matrix_comp} with full reorthogonalization to calculate low-lying eigenvalues and eigenvectors for system sizes up to 
$N_s=169\ (13 \times 13)$ lattice sites.

We also carefully studied the effect of closed and periodic boundary conditions. All the results shown in this paper are obtained with periodic boundary conditions, which preserve the translational 
invariance of the lattice and minimize finite-size effects.

\section{Results: one-dimensional lattice}\label{sec:1D}

In this section we present numerical results for a one-dimensional chain.

For the $n$-th eigenvalue of the system $E_n$, the corresponding eigenstate can be expressed as  
\begin{equation}
\label{wave_vector}
|\psi_n\rangle=\sum_{x_1,x_2,x_3}\psi_n(x_1,x_2,x_3)\hat c_{x_1,1}^\dagger \hat c_{x_2,2}^\dagger \hat c_{x_3,3}^\dagger |0\rangle
\end{equation}
where the same notation as above is used. $|\psi_n(x_1,x_2,x_3)|^2$ provides us with the probability of having the particles with the hyperfine-state $\sigma=1,2,3$ at the positions $x_\sigma$ in 
the $n$-th energy level. Since the system is translationally invariant, we can consider the following 3-particle probability distribution:
\begin{equation}
\label{probability}
\omega_n(x_1,x_2)=\sum_{x_3}|\psi_n(x_1+x_3,x_2+x_3,x_3)|^2
\end{equation}
which is only a function of the relative positions of the fermions with the hyperfine-state $1$ and $2$ with respect to the fermion with hyperfine-state $3$. We can anticipate that our results 
will prove that only the long-distance behavior of the many-body wave-function (and therefore of $\omega$) provides clear-cut signatures of bound-states formation. This is a crucial difference 
between the results presented in the next subsection and Ref. [\onlinecite{Klingschat}], where only the energy spectrum and onsite observables were investigated, and also the main reason we arrive 
at somewhat different conclusions.

\subsection{Lossless regime, $V=0$}\label{subsec:1D_without_constraint}

We start by considering the case where three-body losses do not play any role, i.e. we take $V=0$ in Eq. (\ref{Hamiltonian}).

First we study the evolution of the ground state properties of the system as a function of the interaction.
As shown in Fig.~\ref{GS_wave_function}, for all values of the interaction the ground-state probability distribution $\omega_0(x_1,x_2)$ has a pronounced maximum at $x_1=x_2=0$ and a fast decay with 
the distance away from this point. We also found that, already for a relatively small system size, the tails of the probability distribution at large distances are essentially unaffected by the size 
of the lattice.  Their fast decay is consistent with the formation of three-body bound states (trions) for arbitrarily small interactions. Therefore, similar to the situation for for two-body bound 
states in continuum space \cite{2-body_threshold}, there is no threshold for three-body bound states formation in $1D$. It is easy to realize that at weak-coupling their wavefunction has a finite 
spread in space over many lattice sites and therefore trions can be seen as local objects only for large values of the interaction.   

\begin{figure}[hbpt]
\includegraphics[width=0.45\textwidth]{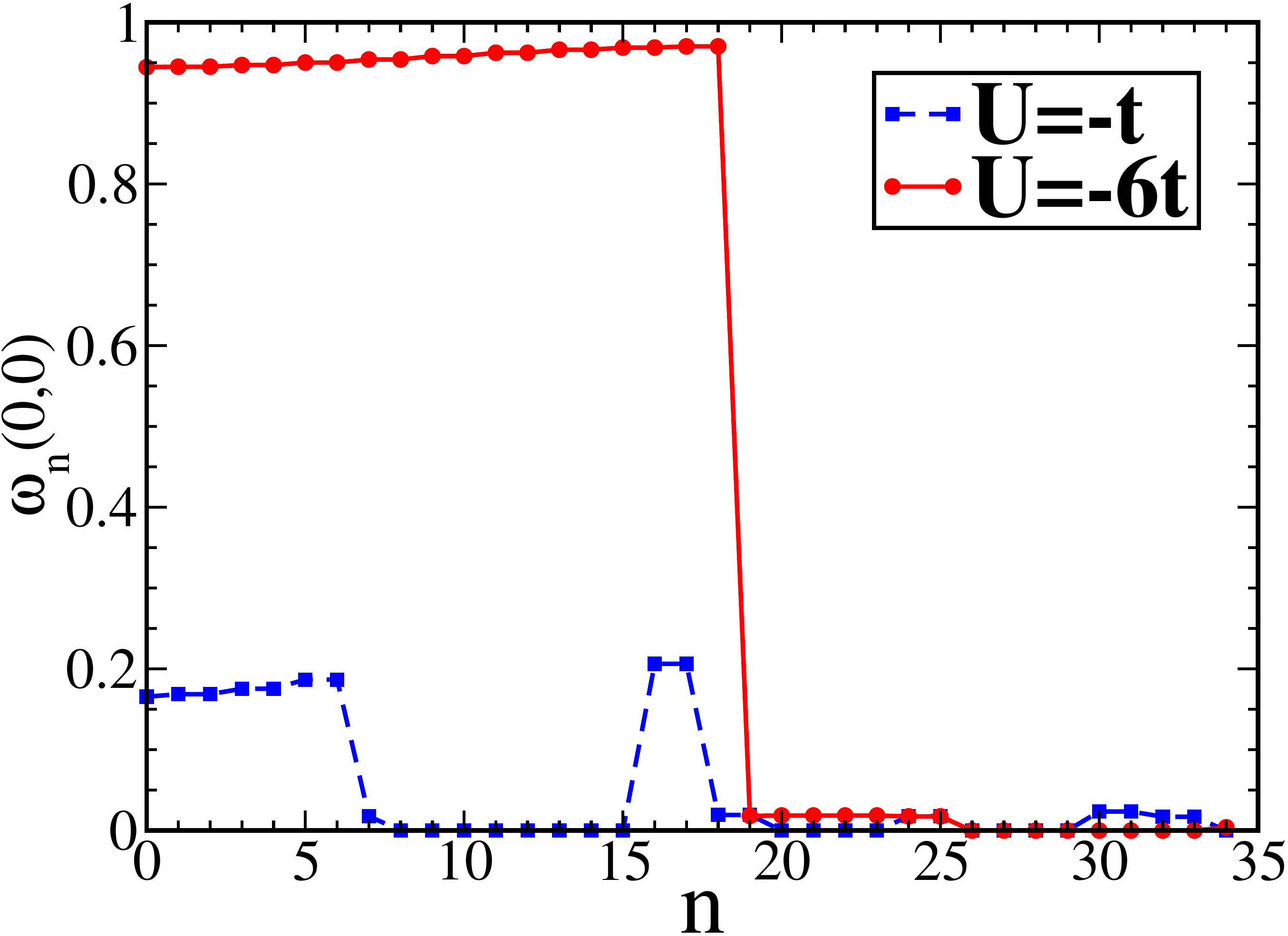}  
\vspace{-0.2cm}
\caption{(Color online) Probability of triple occupation $\omega_n(0,0)$ as a function of the eigenstate label $n$ for increasing energy and different values of the interaction:  $U=-t$ blue dashed line 
with squares, $U/t=-6$ red solid line with circles. The data shown correspond to a linear chain of $N_s=19$}
\label{trionic_weight}
\end{figure}
\begin{figure}[b]
\includegraphics[width=0.45\textwidth]{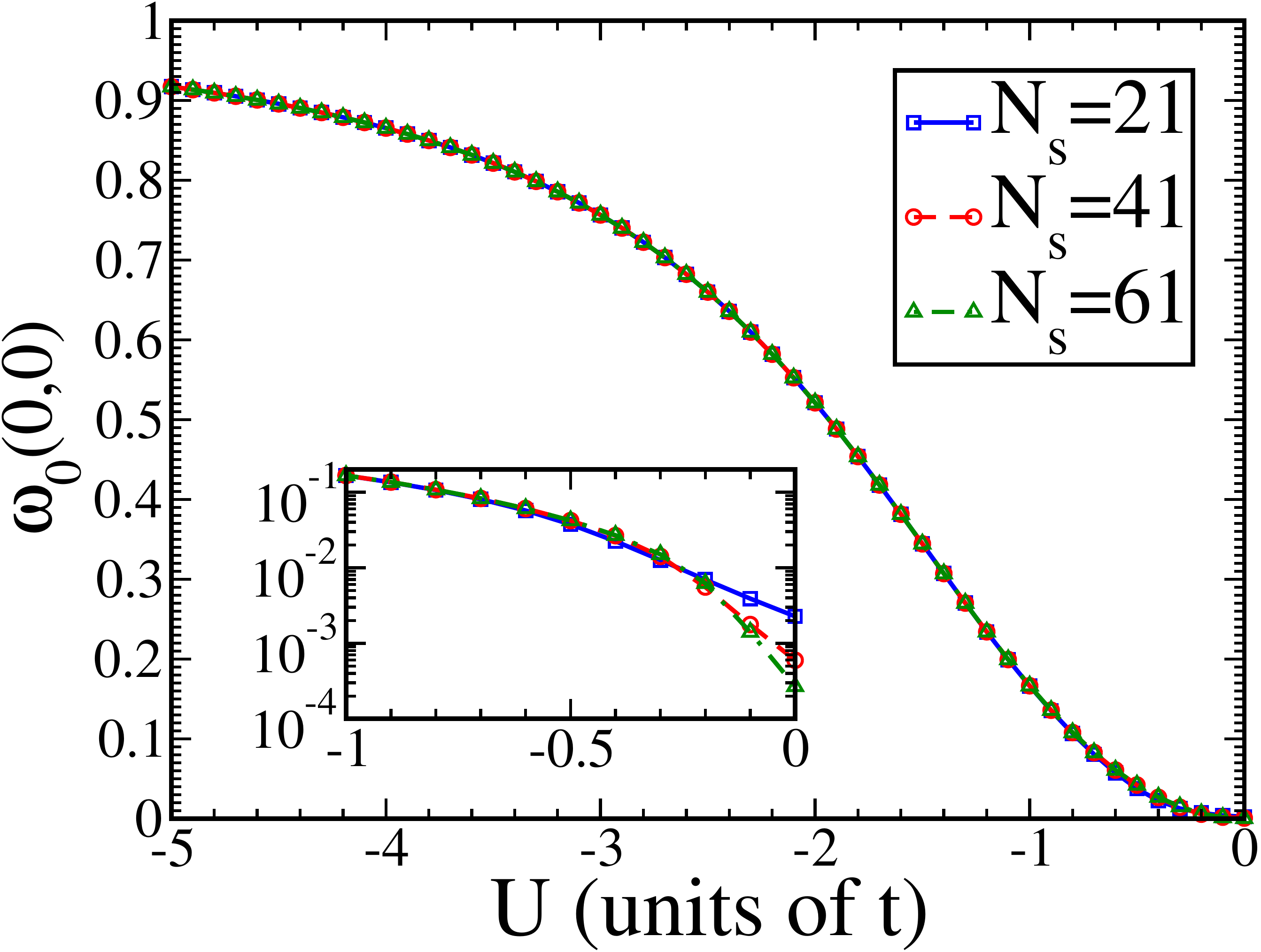}  
\vspace{-0.2cm}
\caption{(Color online) Ground-state probability of triple occupancy $\omega_0(0,0)$ plotted as a function of the interaction for different values of 
the system size: $N_s=21,31,41$ (blue solid line with squares, red dashed line with circles and green dashed-dotted line with triangles respectively).
The inset shows the same plot in log-scale (y-axis), in order to underline the relevance of finite-size effects at weak-coupling.}
\label{GS_trionic_weight}
\end{figure}

\begin{figure}[hbpt]
\begin{center}
\subfigure[ ]{
\label{dispersion-on-site}
\begin{minipage}[b]{0.45\textwidth}
\centering \includegraphics[width=1\textwidth]{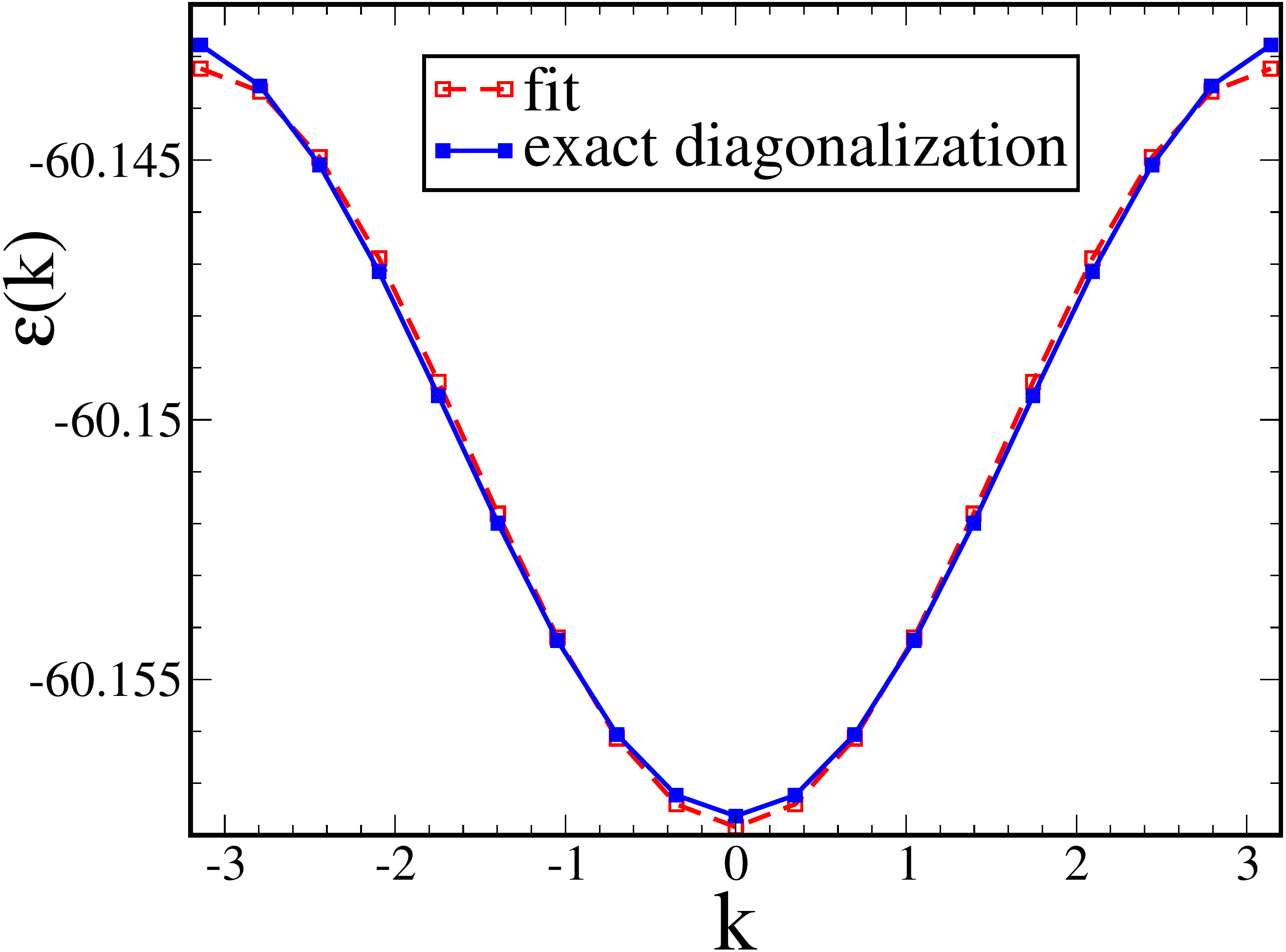}
\end{minipage}}\\
\subfigure[ ]{
\label{dispersion-off-site}
\begin{minipage}[b]{0.45\textwidth}
\centering \includegraphics[width=1\textwidth]{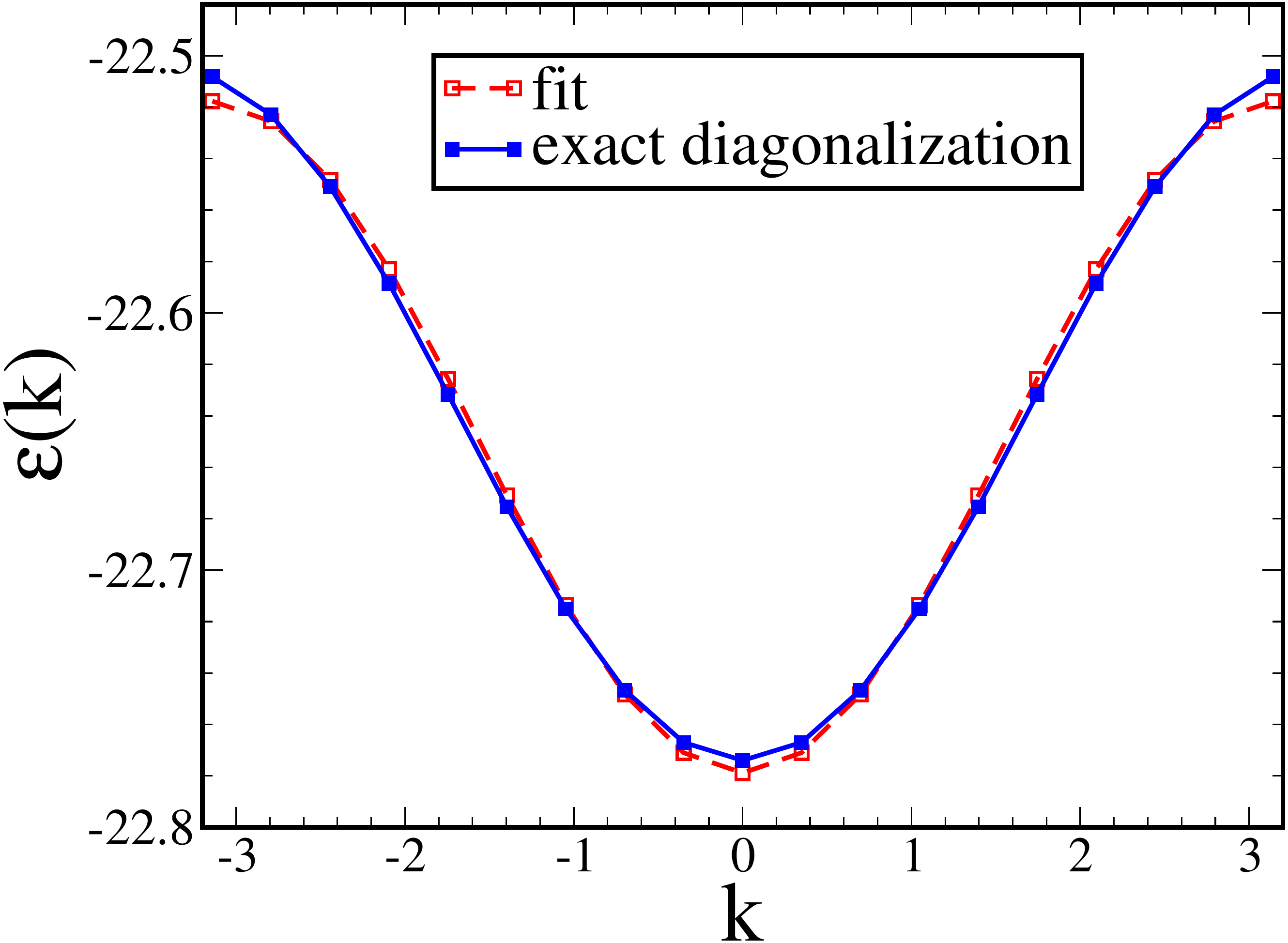}
\end{minipage}}\\
\caption{(Color online)  Dispersion relation for (a) on-site and (b) off-site trionic bands for a lattice with $N_s=19$ and $U=-20t$. The dispersion 
relations for both on-site and off-site trions are well described by the free particle dispersion $\varepsilon(k)=-2J_{eff}\cos(kl) -\mu_{\rm{eff}}$, where $l$ is the lattice spacing assumed as length unit. 
For on-site trions the results are in good agreement with the perturbative result $J_{\rm{eff}}=3t^3/2U^2$ and  $\mu_{\rm{eff}}=-3U-3t^2/U$~[\onlinecite{Titvinidze-Privitera}], as shown by fitting the data
with the asymptotic power-law behavior ($J_{\rm{eff}} \approx 1.46(t^3/U^2)$ and $\mu_{\rm{eff}} \approx -3U-3t^2/U$). For off-site trions instead $J_{\rm{eff}} \approx 1.31 \cdot 10^{-2} (t^2/U)$ 
and $\mu_{\rm{eff}} \approx -2U-2t-13t^2/U$ for this value of $U$.}
\label{dispersion}
\end{center}
\end{figure}

\begin{figure}[hbpt]
\centering\includegraphics[width=0.45\textwidth]{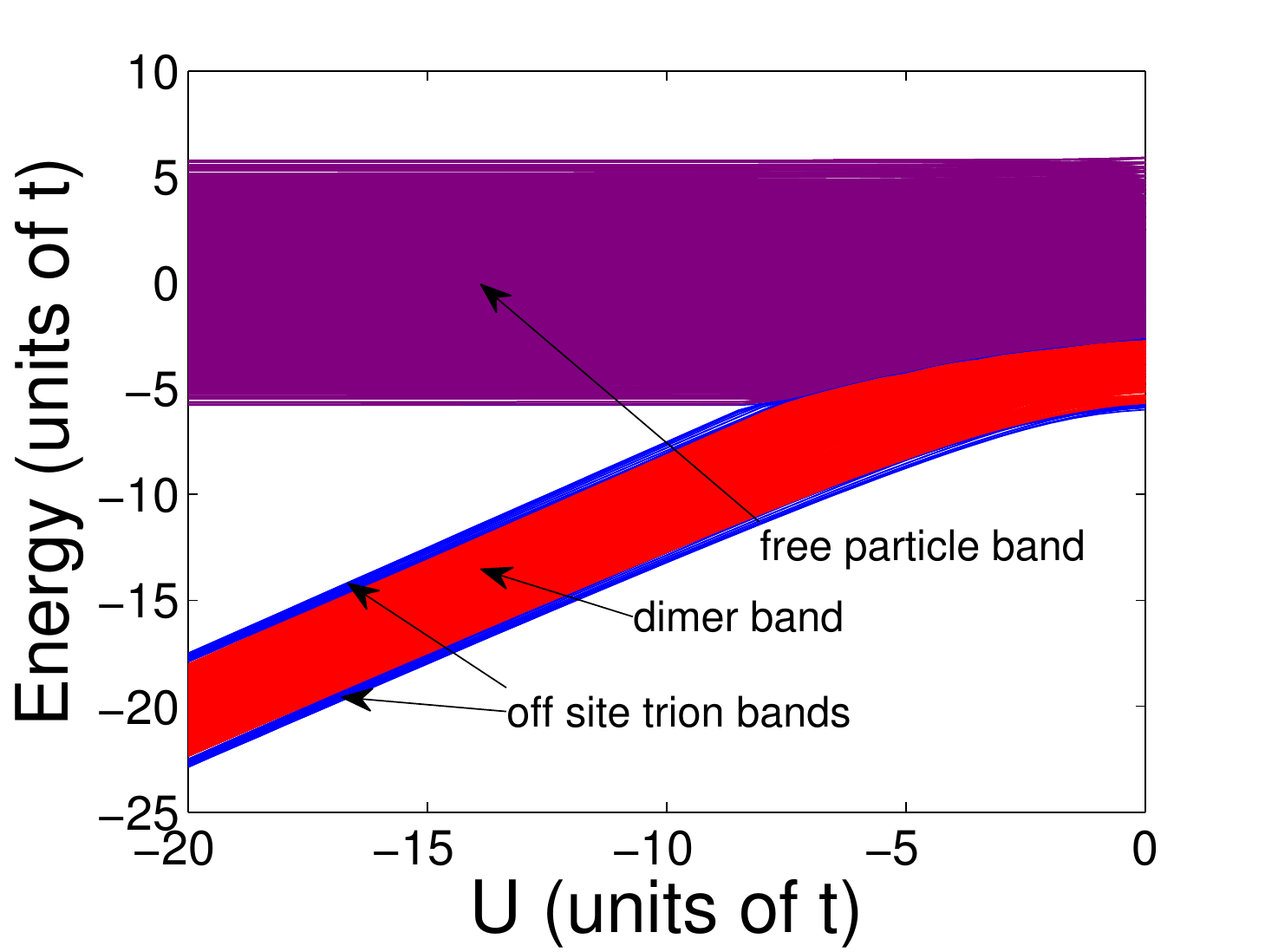}  
\vspace{-0.2cm}
\caption{(Color online) Energy spectrum in 1D including the three-body constraint ($V/t=10^3$) as a function of $U/t $ for $N_s = 19$. 
Different bands are plotted with different color (the off-site trion band is blue, the dimer band is red and 
the free particle band is purple). For weak interactions bands are not separated and therefore colors are only formal.}
\label{Spectrum_1D_constraint}
\end{figure}

The eigenvalue spectrum shown in Fig.~\ref{1D_ns19_spectrum} allows to distinguish three main structures which are clearly separated from each other only for strong interactions, while they overlap at 
weak-coupling. We found that the lowest band separates from the second band for $|U|=U_0=1.78t$ in agreement with Ref. [\onlinecite{Klingschat}].  Moreover a further study of the low-energy spectrum using 
the Lanczos method (see Fig.~\ref{1D_ns51_spectrum}) shows that this threshold converges very quickly as a function of the system size.
      
We then characterize the eigenstates through a detailed study of  $\omega_n$. As one would expect, the analysis reveals that the states of the lowest band (see Fig. \ref{on-site-trions}) are trionic 
in nature and $\omega_n$ is maximal in the center. Moreover, since trions are already essentially local objects for $|U|=U_0$, this provides a simple explanation for the observed system size-independence 
of $U_0$.

More surprising is the existence of a fine structure inside the second band, which is not visible on the scale of the plot in Fig.~\ref{1D_ns19_spectrum}. Indeed the second band is further split into three 
sub-bands which are separated by extremely small gaps at strong-coupling of the order of  $t/2$. 
Despite being so close in energy these sub-bands accommodate states which are very different in nature. 
The existence of a fine structure in the spectrum and the asymptotic size of the gap are in agreement with 
the results found at strong-coupling for the analogous problem with three indistinguishable bosons\cite{Valiente}, 
which also provides an insightful picture of the strong-coupling regime and further confirms our results.

\begin{figure}[hbpt]
\subfigure[]{
\label{2D_ns4_spectrum}
\begin{minipage}[b]{0.45\textwidth}
\centering \includegraphics[width=1\textwidth]{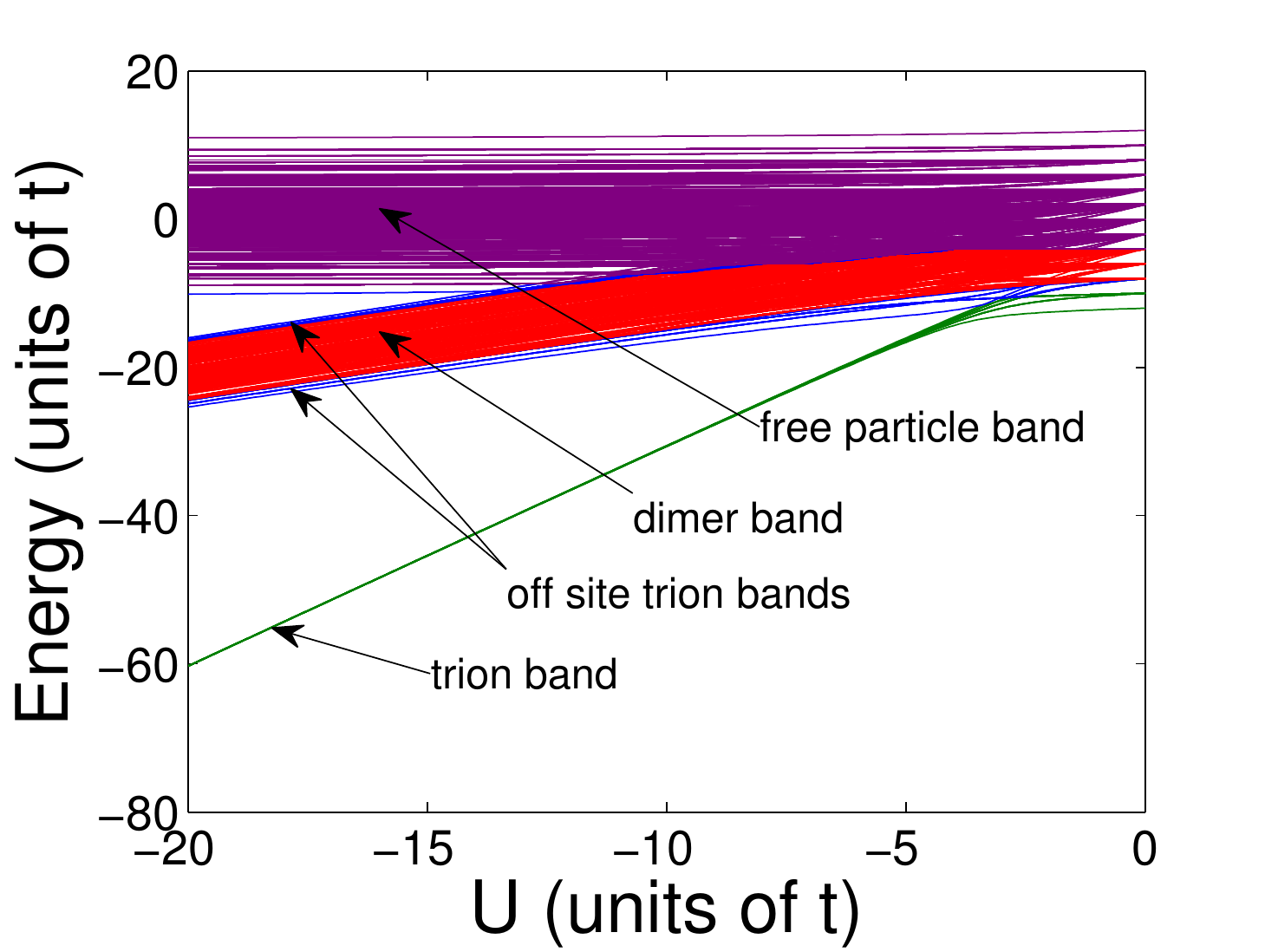}
\end{minipage}}\\
\subfigure[]{
\label{2D_bigger_spectrum}
\begin{minipage}[b]{0.45\textwidth}
\centering \includegraphics[width=1\textwidth]{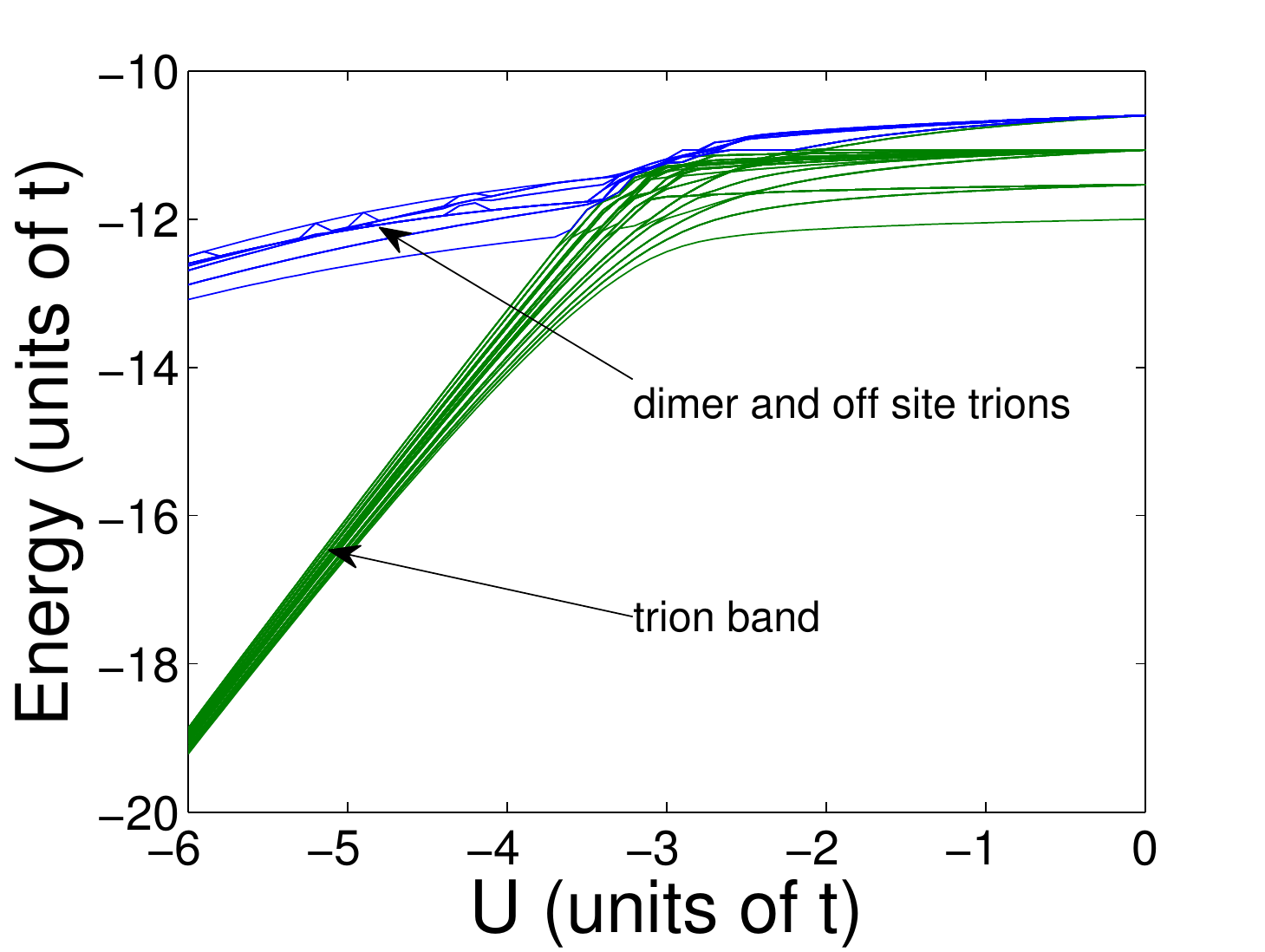}
\end{minipage}}\\
\caption{(Color online) Energy spectrum in 2D as a function of $U/t$. In panel (a) we plot the full spectrum obtained within ED for  $N_s=16 (4 \times 4)$ sites, while in panel (b) we plot 
the first $N=100$ low-lying eigenvalues obtained within the Lanczos method for $N_s=81 (9 \times 9)$. 
Different bands are plotted with different color (the trion band is green, the off-site trion band is blue, the dimer band is red and 
the free particle band is purple). For weak interactions bands are not separated and therefore colors are only formal.} 
\label{Spectrum_2D}
\end{figure}

The central sub-band is composed of proper dimer states where the three-body bound states are dissociated. This is clearly evident in  Figs. \ref{dimer1} and \ref{dimer2}, where $\omega_n$ corresponding to  
different states in the dimer sub-band are shown.  In these cases $\omega_n(x_1,x_2)$ does not decay anymore at large distances and is maximal along the lines $x_1=0$, $x_2=0$ or $x_1=x_2$, being very small in 
the origin $x_1=x_2=0$. These eigenstates correspond to configurations where a two-body bound state coexists with an unbound particle. 

The states below and above the dimer sub-band show instead a completely different behavior, since their  probability distribution has a fast decay at large distances, as shown in Figs. \ref{off-site-trions1} and 
\ref{off-site-trions2}. This indicates that they are excited three-body bound states. This indicates that they are excited three-body bound state, i.e. the third particle still stays bound to a dimer, although 
their energies are very close to dimer states. The probability distribution at $x_1=x_2=0$ is however substantially suppressed as in the dimer states,  while is maximal when a pair sits in one of the  
nearest-neighbors sites of the third particle. We therefore refer to these states as off-site trions, in order to distinguish them from the states in the lowest band. For distinction we will 
refer from now on to the latter as on-site trions. 

Our interpretation is that in the case of off-site trions the increase of local quantum fluctuations with increasing energy is strong enough to reduce the probability of finding three fermions on the same 
site $\omega_n(0,0)$ basically to zero even at strong-coupling, while the three-body bound state character, which is related to the large-distance behavior of $\omega_n$, is not changed. Indeed the probability 
distribution of dimer states and off-site trions are pretty similar at short distances, and this probably also explains why they are so close in energy. We were not able to find simple arguments to explain 
the existence of off-site trions and also the size of the gaps between the off-site trionic and dimer sub-bands, which are for example absent in $D=2$, as shown in Sec. \ref{sec:2D}.

Finally none of the states of the highest band in Fig. \ref{1D_ns19_spectrum} are bound, and the probability distribution $\omega_n$ is essentially flat and small over the whole system, as evident in 
Fig. \ref{particle-band}. We refer to this structure as the unbound particles band.  

It is thus interesting to observe that having a maximum in the probability of finding three-particles in the same lattice site is not essential to three-body bound state formation, which is instead unambiguously 
related to the long distance behavior of the probability distribution. 

In order to better understand this point we further studied $\omega_n(0,0)$,  i.e. the probability to find all three particles at the same lattice site. This parameter was used in Ref. \onlinecite{Klingschat} 
in order to identify trions and hence called {\it{trionic weight}} by the authors. From the considerations above it is however clear that the existence of off-site trions and the non-local character of 
the trionic wave-function at weak-coupling makes this interpretation inconclusive. 

Indeed,  the very irregular behavior of $\omega_n(0,0)$ shown in Fig. \ref{trionic_weight} in the weak-coupling regime is simply due to the fact that the onsite trionic band overlaps with the higher bands. 
Therefore  some of the states shown in Fig. \ref{trionic_weight} are onsite trions, while others are still trions but off-site in character or just dimers. For small $U$ these states are very close in energy 
and only the detailed study of the eigenstates that we performed can distinguish between them. Even though the states in Fig. \ref{trionic_weight} with a large $\omega_n(0,0)$ are indeed on-site trions, 
a decrease in $\omega_n(0,0)$ does not always correspond to an unbound state because of the presence of off-site trions. Moreover, in the weak-coupling regime there are dimer or even specific unbound 
particle states which have a probability for triple occupation even larger than for three-body bound states.

\begin{figure}[htb]
\subfigure[]{
\label{On-site-trions_2D_a}
\begin{minipage}[b]{0.45\textwidth}
\centering \includegraphics[width=1\textwidth]{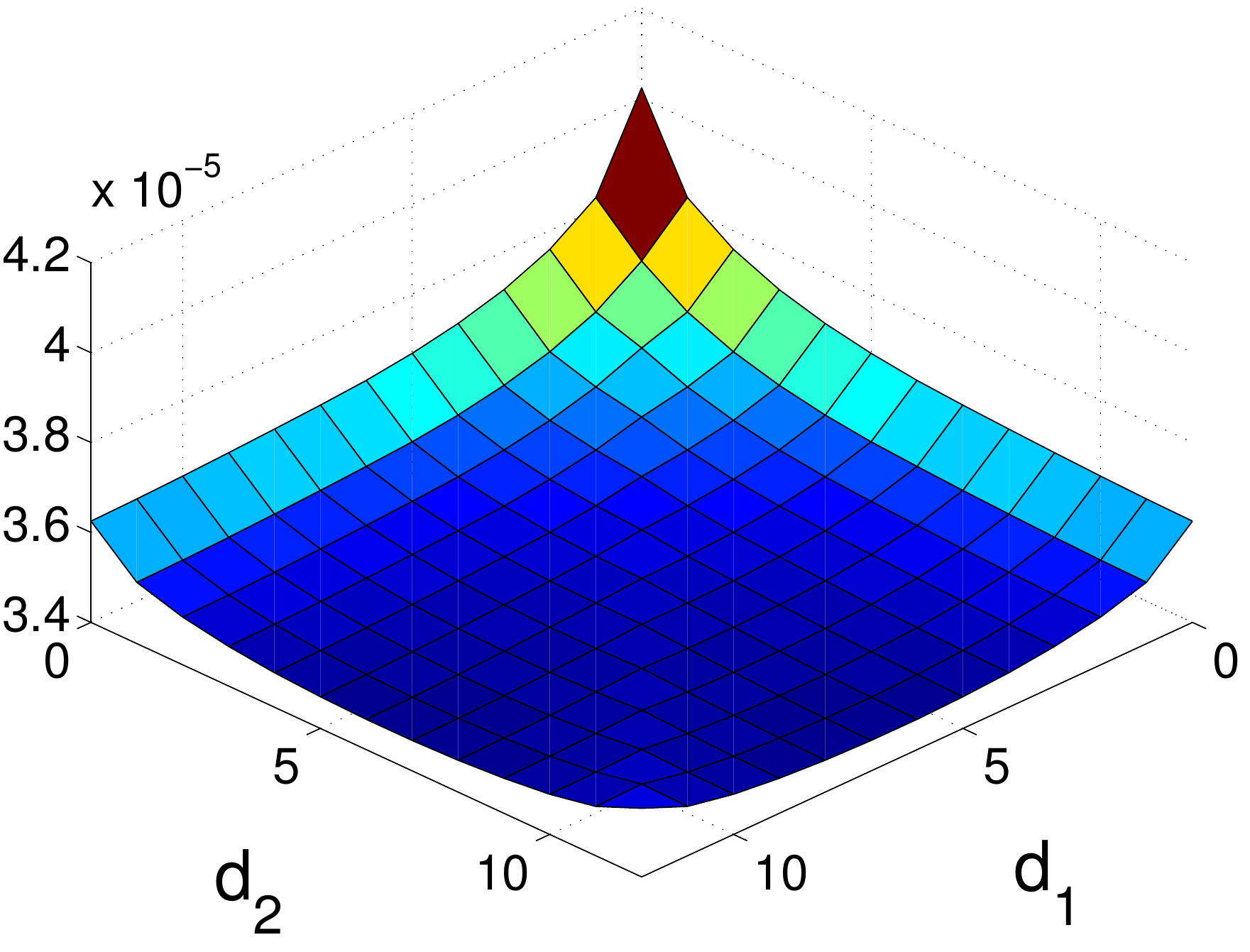}
\end{minipage}}\\
\subfigure[]{
\label{On-site-trions_2D_b}
\begin{minipage}[b]{0.45\textwidth}
\centering \includegraphics[width=1\textwidth]{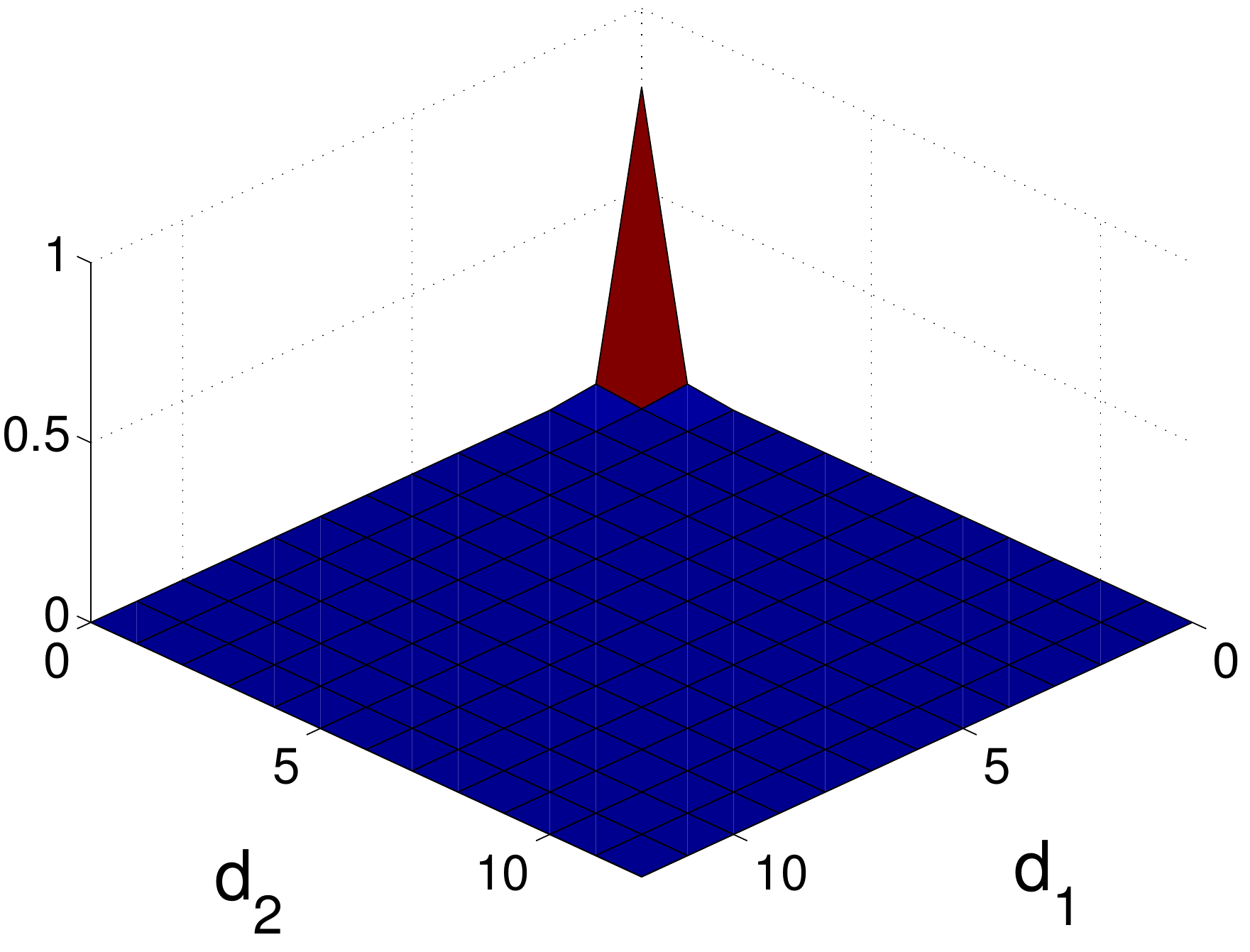}
\end{minipage}}\\
\caption{(Color online) Ground-state  probability  distribution $\bar\omega_0^{2D}(d_1,d_2)$  for a 
 a 2D square lattice with  $N_s= 169 (13 \times 13)$: (a)  weak-coupling ($U/t=-0.1$), and (b) strong-coupling ($U/t=-4.8$) as a function of the Manhattan distances $d_1$ and $d_2$ (see main text).  }
\end{figure}

Our point made above is dramatically evident at strong coupling, where the onsite trionic band is clearly separated in energy from the higher bands. Indeed, while $\omega_n(0,0)$ is close to one for all 
the states in the onsite trionic band, indicating that onsite trions are essentially local objects for this value of the coupling, the same parameter drops to a very small value for off-site trions, 
which are still three-body bound states.  Within the onsite trionic band, we also observe as in Ref. [\onlinecite{Klingschat}] that excited trions have a slightly higher value of $\omega_n(0,0)$, compared 
to the ground state. This is probably related to the higher potential energy contribution in the excited states and to the different relevance of quantum fluctuations.

In conclusions local observables like $\omega_n(0,0)$ are unable to distinguish between an off-site trionic and  dimer states in all interaction regimes although their wave-functions are completely different, 
as shown above. Therefore $\omega_n(0,0)$ should in general not be taken as an indicator of three-body bound state formation.

As shown in Fig. \ref{GS_trionic_weight},  we found the parameter $\omega_0(0,0)$ in the ground state to increase smoothly with $|U|$ independently of the system size in contrast 
to Ref. [\onlinecite{Klingschat}], where a step-like behavior was observed. 

In the strong-coupling limit we expect that trions behave as composite particles and it makes sense to investigate their dispersion relation. The results shown in Fig. \ref{dispersion} prove that 
trionic states are distributed in their band as free-particles in the tight-binding approximation. For onsite trions the  effective chemical potential $\mu_{\rm{eff}}$ and hopping $J_{\rm{eff}}$ 
agree very well with perturbative estimates given in Ref. [\onlinecite{Titvinidze-Privitera}]. For off-site trions instead $J_{\rm{eff}} \propto t^2/U$

\begin{figure}[b]
\begin{center}
\includegraphics[width=0.45\textwidth]{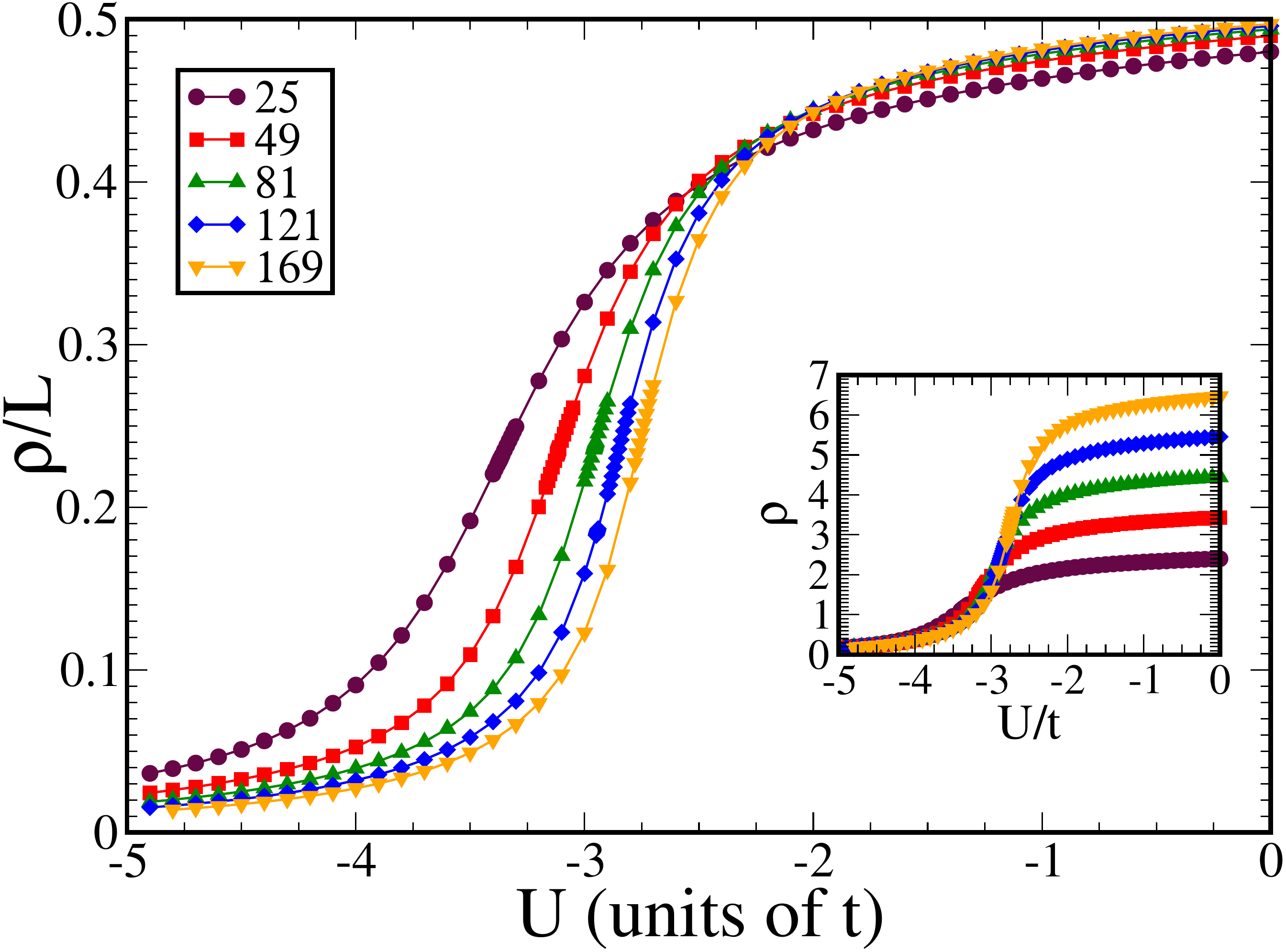}
\caption{(Color Online) Average size $\rho$ of the three particle system as a function of  $U/t$ 
for different values of lattice size $N_s=25,49,81,121,169$. The inset shows the ratio $\rho/L$, where $L$ is linear size of the system.}
\label{average_size}
\end{center}
\end{figure}
 

\begin{figure*}[t!]
\subfigure[exited on-site trion]{
\label{Exited_On-site-trions_2D}
\begin{minipage}[b]{0.32\textwidth}
\centering \includegraphics[width=1\textwidth]{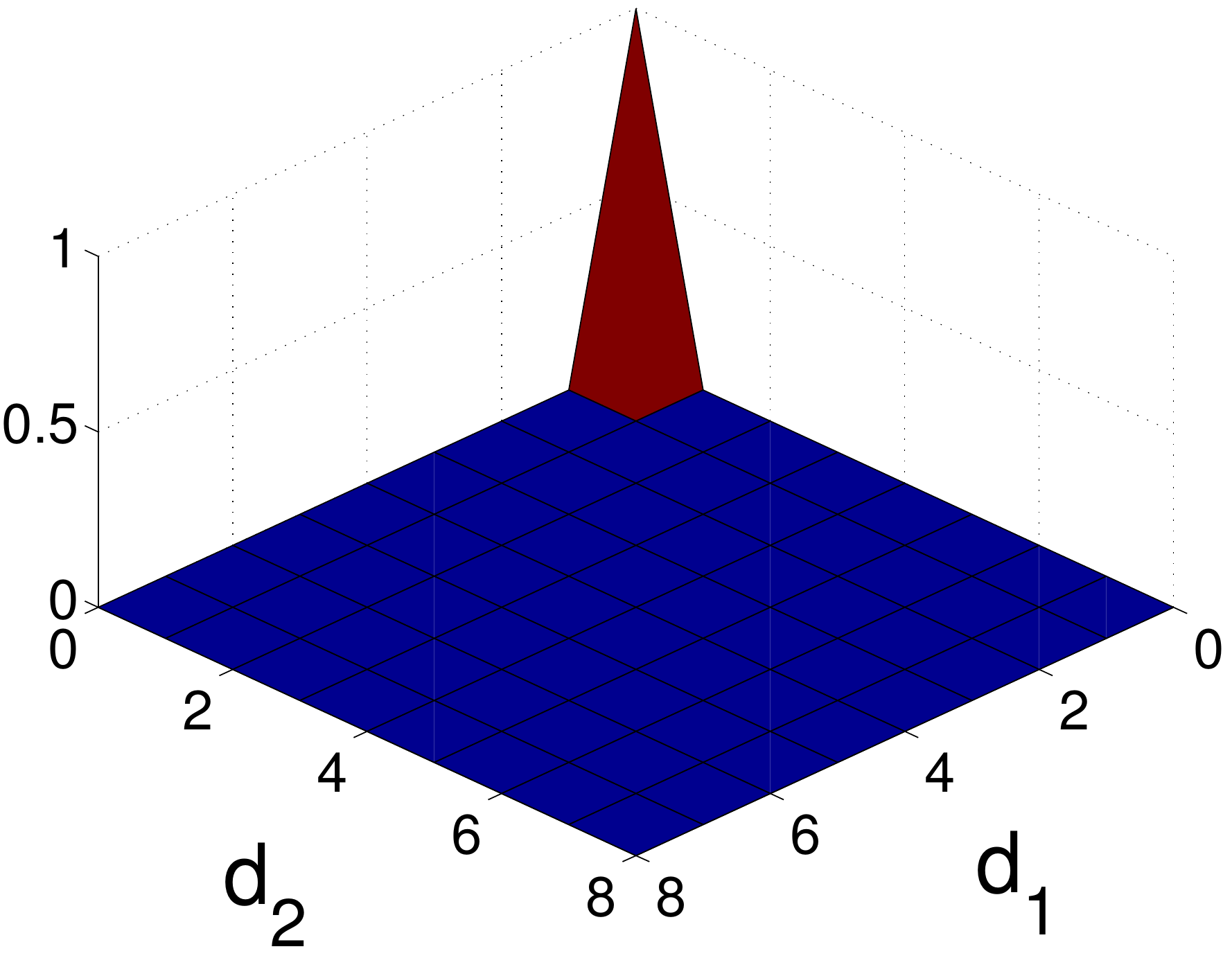}
\end{minipage}}
\subfigure[off-site trion]{
\label{Off-site-trions_2D}
\begin{minipage}[b]{0.32\textwidth}
\centering \includegraphics[width=1\textwidth]{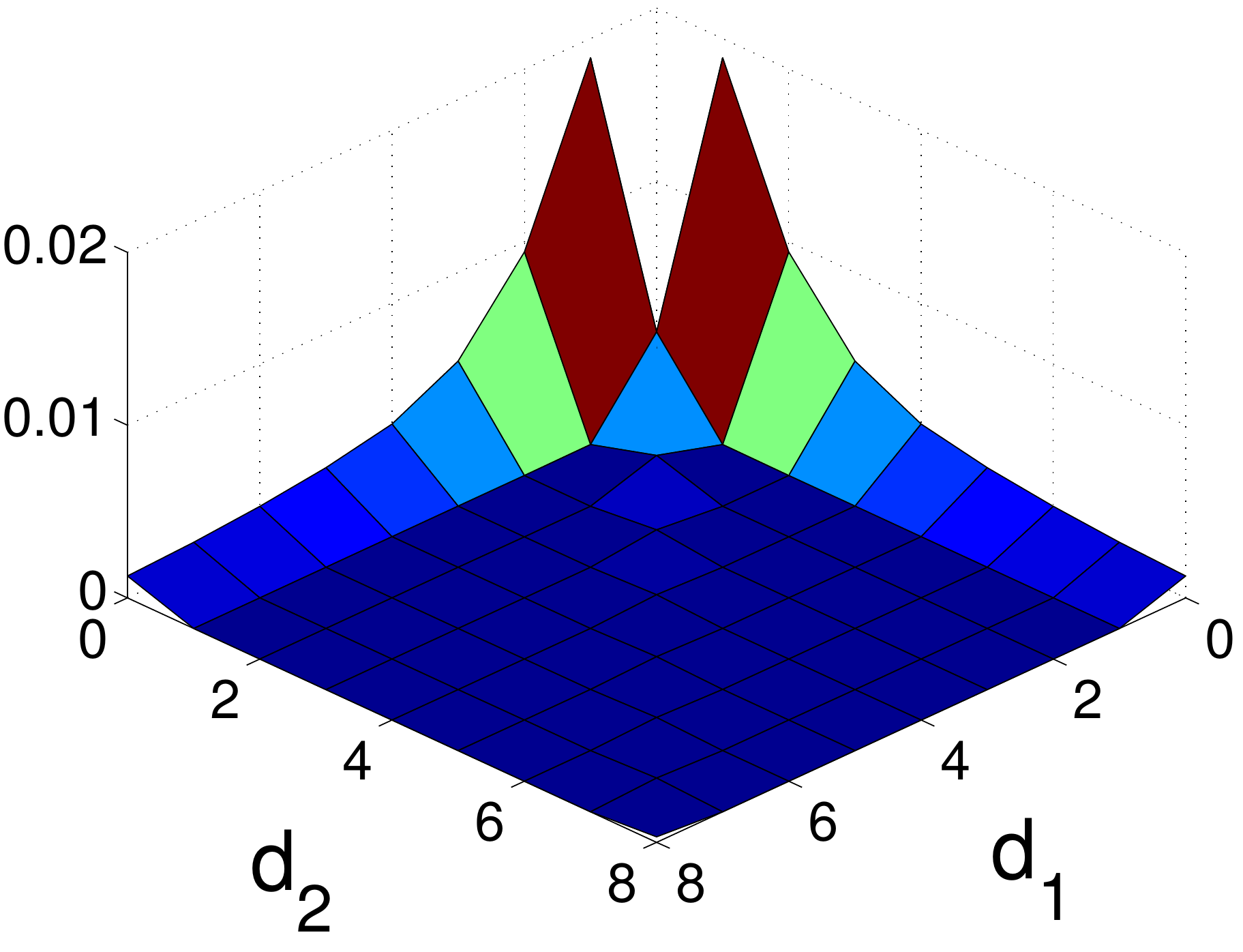}
\end{minipage}}
\subfigure[dimer]{
\label{Dimer_2D}
\begin{minipage}[b]{0.32\textwidth}
\centering \includegraphics[width=1\textwidth]{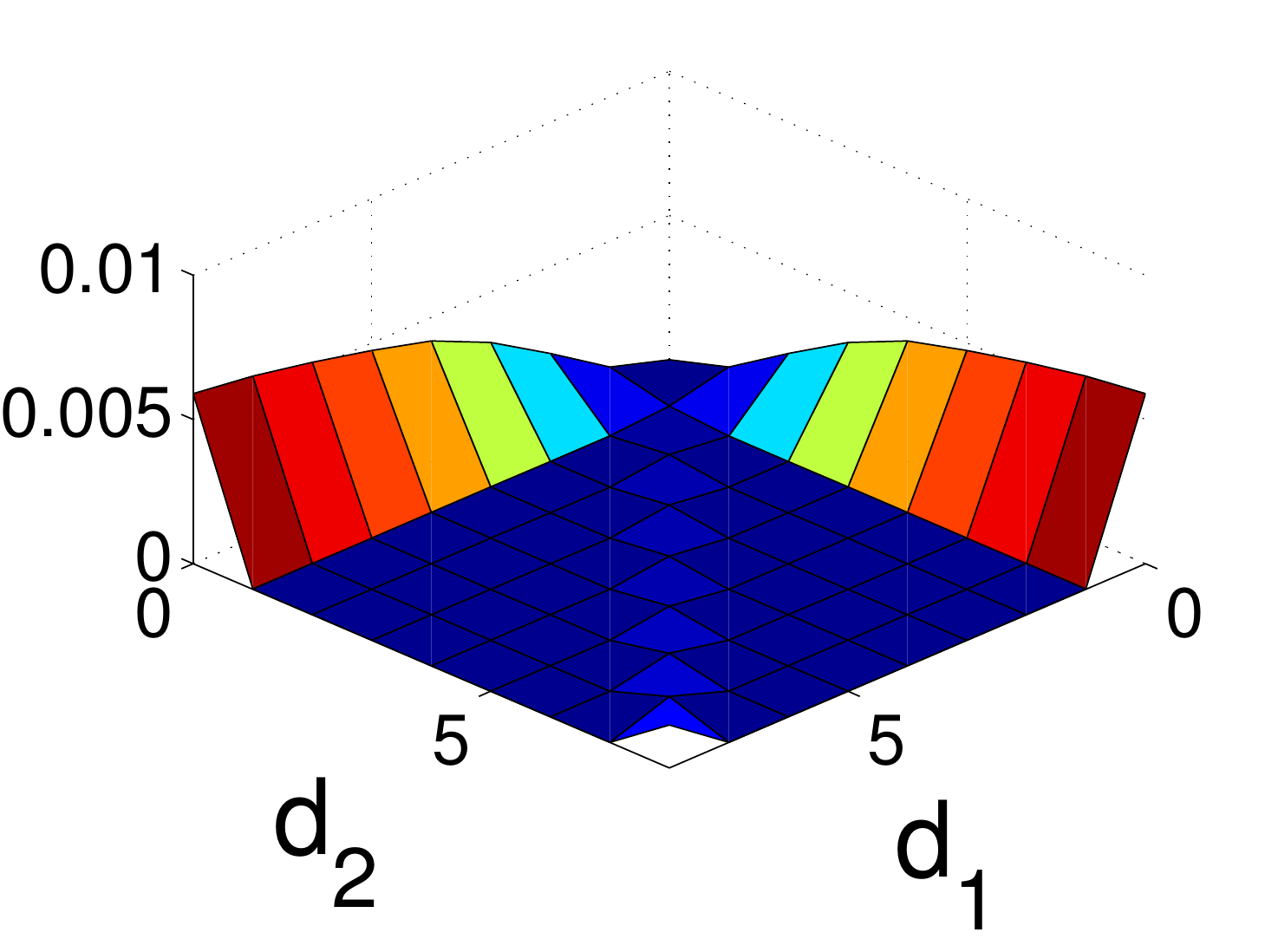}
\end{minipage}}\\
\caption{(Color online) Probability distribution $\bar\omega_n^{2D}(d_1,d_2)$ in a 2D square lattice with $N_s=81 (9 \times 9)$  corresponding to (a) an excited on-site trion, (b) an off-site trion, and (c) a dimer. The interaction is again chosen to be deep in the strong-coupling regime ($U/t=-50$) where the different bands are well separated.}
\label{Exited_2D}
\end{figure*}

\subsection{Strong-loss regime, $V=\infty$}\label{subsec:1D_with_constraint}

As already mentioned in the introduction, realizations of three-component Fermi system in ultracold atomic gases are plagued by three-body losses, which strongly limit the lifetime for observing 
the interesting phenomena under discussion here. However, as shown in Refs.~[\onlinecite{Daley,Kantian}] in the presence of a strong loss rate $\gamma_3/t \gg 1$ in a lattice, actual losses are 
suppressed and the system can be described by the Hamiltonian in Eq. \ref{Hamiltonian} provided one includes an effective three-body constraint ($V \to \infty$). Therefore, in order to understand 
the effect of a strong three-body loss rate on three-body bound state formation, we include the additional three-body term in the Hamiltonian and set $V/t=10^3$. As proven in 
Ref.~[\onlinecite{Titvinidze-Privitera}], this setup is indeed essentially indistinguishable from the real constraint. 
 
The energy spectrum of the system in the presence of the three-body constraint is shown in Fig. \ref{Spectrum_1D_constraint}.
Due to the constraint, triple occupancies are forbidden and therefore onsite trions are suppressed (i.e. their energy is lifted by an amount proportional to $V$). 

One could then wonder if an onsite three-body constraint  is sufficient to rule out the possibility of three-body bound states in the presence of an attractive two-body potential which is also onsite. 
The previous considerations on the existence of off-site trions clearly indicate that the probability of having triply occupied sites is not necessarily correlated with the formation of three-body bound states. 
Our results fully confirm this expectation since in the constrained case we found the ground state of the system to be always a three-body bound state but off-site in nature. As before the off-site trionic 
states are separated by a small energy gap from the dimer states. Since both off-site trionic and dimer states have similar values of $\omega_n(0,0)$ already in the unconstrained case, the energetic hierarchy 
between them is only weakly affected by the constraint, despite of the suppression of triple occupancies. The existence of off-site trions and hence of an off-site trionic phase has been also observed 
in Ref.~[\onlinecite{Kantian}],  where the many-body version of the problem investigated here was addressed in one spatial dimension.  

Therefore, at least in 1D, three-body bound state formation is a robust phenomenon which takes place for arbitrarily small values of the interaction and is robust against three-body losses. Moreover, 
the presence of off-site trions show that quantum fluctuations can strongly suppress triple occupancies without necessarily inducing the breaking of trions in favor of dimers. In order to understand 
how much these conclusions are related to the peculiarity of the one-dimensional case, in the next section we will address the same problem in the context of a two-dimensional square lattice.

\section{RESULTS: TWO-DIMENSIONAL LATTICE}\label{sec:2D}

As in the one-dimensional case, we first investigate the case where no losses are present and $V=0$. 
The complete energy spectrum of the system on a square lattice with $N_s=16 \ (4 \times 4)$  lattice site as a function of $U$ is shown in Fig. \ref{2D_ns4_spectrum}. 
Despite the same overall structure as in the one-dimensional case emerges from the picture, finite-size effects are clearly not-negligible. The fast growth of the Hilbert space with the number $N_s$ of 
lattice sites strongly constrains the linear size of the lattice which we can handle within full ED and even within the Lanczos algorithm we were only able to address up to $N_s=169\ (13\times 13)$ sites 
(see Fig. \ref{2D_bigger_spectrum}). Since finite-size effects are expected to scale with the dimension $D$ as a function of $N_s^{-1/D}$, it is clear that increasing dimensionality is extremely 
unfavorable for our numerical approach and a careful extrapolation to the thermodynamic limit is required for all observables.   

As in 1D the lowest band separates from the higher bands only beyond a critical coupling. The threshold value is roughly twice larger than the value in 1D ($|U_0^{2D}|/t \approx 3.4$ for $N_s=16$), 
suggesting that they could converge to the same value when rescaled with the bandwidth $W \propto D$. The prefactor is however slightly bigger ($|U_0^{2D}|/t \approx 3.6$) in the data obtained within 
the Lanczos approach on a lattice with $N_s=169$ shown in Fig. \ref{2D_bigger_spectrum}, suggesting that there could still be sizable finite-size effects affecting the threshold value.

It is worth pointing out that the lowest band now follows rather closely the higher bands in a finite range of couplings, as evident in Fig. \ref{2D_bigger_spectrum}. Since this could suggest the existence 
of a finite threshold for three-body bound state formation which is absent in 1D, we also studied in detail the eigenvectors and the probability distribution 
$\left|\psi\left(x_1,y_1,x_2,y_2,x_3,y_3\right)\right|^2$ in the two-dimensional case.  

By using again the translational invariance, we  obtain the following  reduced probability distribution:
\begin{eqnarray}
\label{probability_2D_0}
&&\omega_n^{2D}(x_1,y_1,x_2,y_2)\\
&&=\sum_{x_3}|\psi_n(x_1+x_3,y_1+y_3,x_2+x_3,y_2+y_3,x_3,y_3)|^2 \nonumber \,.
\end{eqnarray}

Since, however,  $\omega_n^{2D}$ is still a function of four variables, we further reduce them by averaging the probability distribution $\omega_n^{2D}(x_1,y_1,x_2,y_2)$ over all sites at 
the same Manhattan distance\cite{Manhattan} and define 
\begin{equation}
\label{probability_2D}
\bar \omega_n^{2D}(d_1,d_2)\\
=\sum \omega_n^{2D}(x_1,y_1,x_2,y_2)\frac{\delta_{|x_1|+|y_1|,d_1}\delta_{|x_2|+|y_2|,d_2}}{C_{d_1} C_{d_2}}
\end{equation}
In the expression above the summation runs over $-L/2<x_1,y_1,x_2,y_2\leq L/2$ and $C_d$ is the number of the lattice sites at a Manhattan distance $d$ from the third-species fermion. 
While this allows us to produce meaningful three-dimensional plots also in 2D, it clearly averages out a part of the information contained in $\omega_n^{2D}$. 

Moreover, we introduce the following quantity 
\begin{equation}
\label{average_distance}
\rho=\sum_{d_1,d_2} d_1 C_{d_1} C_{d_2}\bar\omega_{2D}(d_1,d_2) \, .
\end{equation}
which measures the average diameter of the three-particle cluster, as clarified below. 

We first consider the evolution of ground state properties as a function of the interaction. 
In the strong-coupling regime shown in Fig. \ref{On-site-trions_2D_b}, where the bands 
in the spectrum are well separated, the probability distribution has a clear maximum at $d_1=d_2=0$ 
and a very fast decay with distance. In this regime the average size of the  cluster $\rho$ plotted in 
Fig.  \ref{average_size} is extremely small and essentially independent of the system size. 
Thus we can confidently argue that at strong-coupling the ground state is an on-site trion. 
	
Due to the relevance of finite-size effects it is much harder to draw analogous conclusions in 
the weak-coupling regime, as shown in \ref{On-site-trions_2D_a}. Indeed in this case the probability 
distribution stays finite at the edges of the lattice even for $N_s=169$, as it was the case for dimer 
states in 1D. On the other hand, since also three-body bound states are expected to have a rather large 
spread of the wavefunction in the weak-coupling regime, the finite value found at the border is equally 
likely to be a residual finite-size effect. 

\begin{figure}[hbpt]
\centering\includegraphics[width=0.45\textwidth]{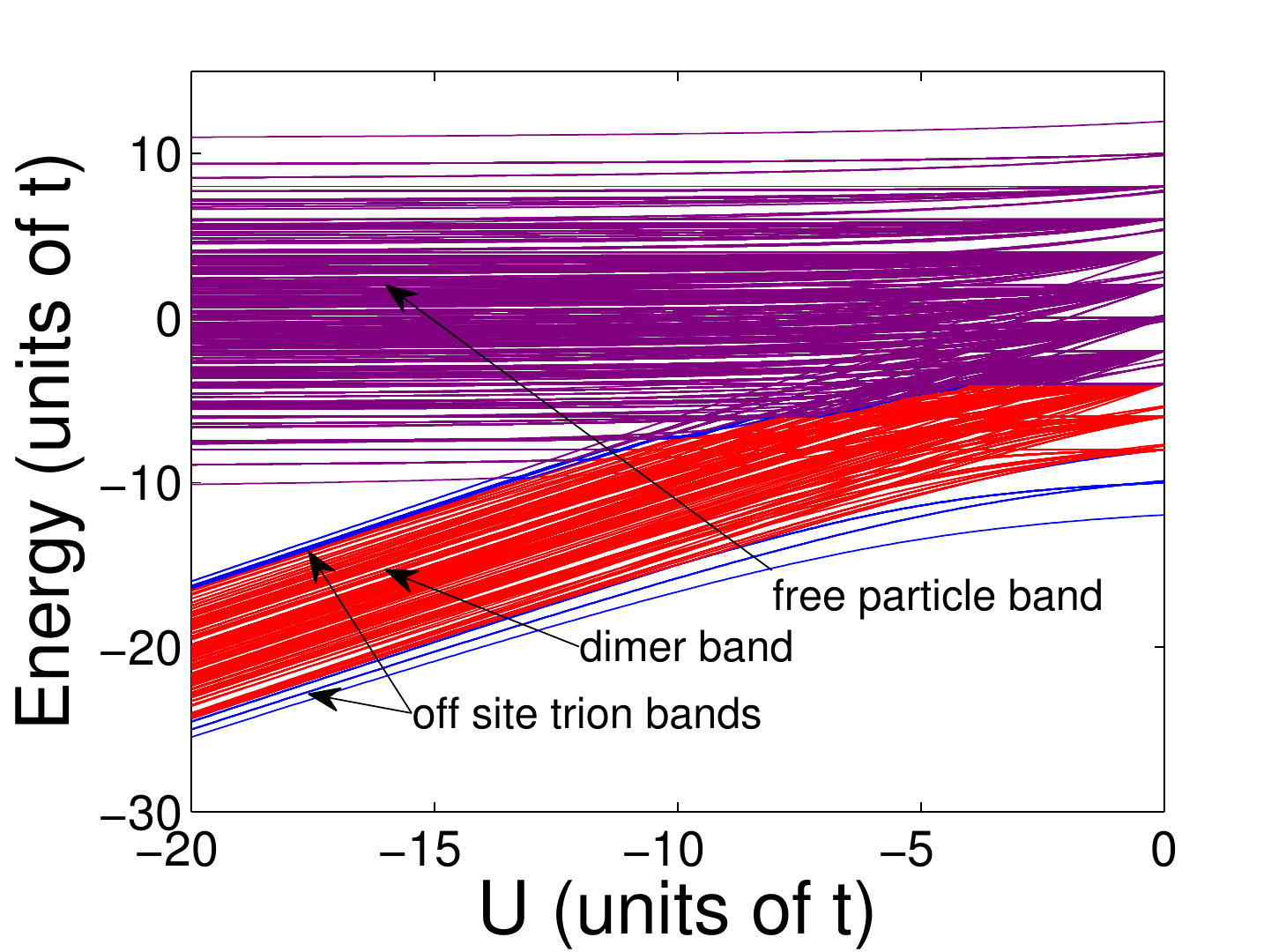}  
\vspace{-0.2cm}
\caption{(Color online) Energy spectrum in 2D including the three-body constraint ($V/t=10^3$) as a function of $U/t $ for $N_s = 16 (4 \times 4)$. 
Different bands are plotted with different color (the off-site trion band is blue, the dimer band is red and 
the free particle band is purple). For weak interactions bands are not separated and therefore colors are only formal.}
\label{spectrum_2D_constraint}
\end{figure}

In order to further clarify this issue we investigated the ratio $\rho/L$ as a function of the interaction strength for different values of $N_s$ 
(see  inset in Fig. \ref{average_size}). One would expect indeed a different scaling with the system size for trions and dimer states. 


Our data clearly show that at strong-coupling $\rho/L$ scales to zero as expected for three-body bound states. 
At weak-coupling instead  $\rho/L$ seems at first-sight to converge to a finite value, with a steep crossover between the two regimes around $|U|=U_c$, 
where  we identify $U_c$ as the inflection point of $\rho(U)$ where $(\partial^2\rho/\partial U^2)_{N_s}=0$. This threshold value clearly decreases in modulus 
with increasing system size, but we are unable to determine whether $U_c$ converges to zero or a finite value in the thermodynamic limit, due to the finiteness of our system. 

It is worth mentioning that $U_c (N_s)$ seems to be compatible with a logarithmic scaling with the system size, as evident from fitting our numerical results. 
While this makes virtually impossible to draw a clear-cut conclusion only based on our exact-diagonalization method, it has instead a relevance for real systems. 
Indeed this would imply the existence in finite 2D  systems, such like e.g. in ultracold gases experiments, of a finite threshold for the crossover to a three-body 
bound state for all the practical purposes.  Similar arguments have already been  used in the context of ultracold gases\cite{Zoran} to explain the existence of 
quasi-condensates in bosonic systems in 2D, in analogy with the existence of a finite magnetization at finite temperature in the finite two-dimensional XY model \cite{Bramwell}. 

The study of excited states resulted in an outcome very similar to the 1D case (see Fig. \ref{Exited_2D}). The main difference compared to the one-dimensional case is that the fine structure within the 
second band in the energy spectrum is now absent and  off-site trionic and dimer states are mixed.  This is probably due to the larger bandwidth in 2D than in the one-dimensional case, which causes the 
different sub-bands to overlap.  The lowest band still accommodates only on-site trions (see Fig.~\ref{Exited_On-site-trions_2D}) and the lowest state in the second band is still an off-site trion 
(see Fig.~\ref{Off-site-trions_2D}), at least in the strong-coupling regime where the nature of the three-body bound state is evident in our data.

Finally we set $V/t=10^3$ and study the effect of the loss-induced constraint on the two-dimensional system. 
The picture which emerges (see Fig. \ref{spectrum_2D_constraint}) is pretty similar to the one-dimensional case and the lowest-energy off-site trion is promoted to the role of ground-state in presence 
of constraint. Due to the greater delocalization of the off-site trionic wavefunction, it is even harder to assess the formation of three-body bound states at weak coupling in this case.

\section{CONCLUSION}\label{sec:Conclusion}
In this work  we addressed the problem of three fermions in different internal states  loaded in a lattice, using an exact diagonalization approach. We studied the attractive and full $SU(3)$-symmetric case, focusing on the competition between three-body and two-body bound state formation. 

Our results show that in one dimension three-body bound states (trions) are formed for arbitrarily 
small attraction between the fermions and they remain the lowest energy states for every value of the interaction. 
We also observed the existence of excited three-body bound states which share the same short-distance structure and typical energy of two-body bound states (dimers),  
while they still feature the characteristic long-distance decay in the wavefunction of three-body bound states. The probability of triple occupancy is strongly suppressed in 
these states, in contrast to the "canonical" trionic states. We therefore refer to them as off-site trions in order to distinguish them from the states in the lowest energy band (on-site trions). 

These results prove unambiguously that the knowledge of the spectrum and onsite observables alone, like in Ref. [\onlinecite{Klingschat}], do not provide a clear-cut distinction among two-body and  
three-body bound-states and can thus lead to misleading conclusions. In this respect only the detailed study of the long-distance behavior of the 3-body eigenstates is able to provide a proper 
characterization of the problem. 

The introduction of a three-body constraint mimicking three-body losses in the strong-loss regime as in Ref.~[\onlinecite{Kantian}], inhibits on-site trions formation and promotes off-site trions to the role 
of lowest energy states. 

In two dimensions, our data suggest a logarithmic scaling with the system size of the threshold interaction for three-body bound states formation. The greater relevance of finite-size 
effects in the two-dimensional case did not allow us to assess  if there is a finite threshold in the thermodynamic limit, although the scaling we found implies the existence of a crossover at a finite 
value of the interaction for all practical purposes in finite systems, like e.g. in ultracold gases experiments. This also suggests that a finite threshold  is likely to exist in the three-dimensional case, 
which we cannot address at present using the methods in this paper and which we plan to address in a future work by using Monte Carlo techniques. 

\begin{acknowledgments} This work was supported by the Deutsche Forschungsgemeinschaft (DFG) via Sonderforschungbereich SFB/TR 49 and Forschergruppe FOR 801. AP thanks C. Verdozzi and the Division of 
Mathematical Physics at Lund University for their hospitality during the completion of this work and M. Capone for the financial support by the European Research Council under FP7/ERC Starting Independent 
Research Grant ``SUPERBAD" (Grant Agreement n. 240524). IT thanks SFB 925, project B5 for financial support.
\end{acknowledgments}



\begin{thebibliography}{99}
\bibitem{Manhattan}
The Manhattan distance is defined as the distance between two points on a square grid, along the grid lines.
\end{thebibliography}
\end{document}